\documentclass{article}
\usepackage[T1]{fontenc}

\usepackage{hyperref,graphicx,amsmath,amssymb}
\usepackage{cite}

\newcommand{\cat}[1]{\left|#1\right\rangle}
\newcommand{\brcat}[2]{\left\langle#1|#2\right\rangle}
\newcommand{\roa}[1]{\left|#1\right\rangle\left\langle#1\right|}
\newcommand{\ro}[2]{\left|#1\right\rangle\left\langle#2\right|}
\newcommand{\aver}[1]{\left\langle #1\right\rangle}
\newcommand{\tr}[1]{Tr{\left( #1\right)}}
\newcommand{\mess}[2]{\left\langle #1\left|#2\right|#1\right\rangle}

\newcommand{\sgn}[1]{^{\left\lbrace #1 \right\rbrace} }
\begin{document}
	
\title{Brief Theory of Multiqubit Measurement}

\author{Constantin Usenko}

\maketitle

\begin{abstract}
	
	Peculiarities of multiqubit  measurement are for the most part similar to peculiarities of measurement for qudit  -- quantum object with finite-dimensional Hilbert space.

	Three different interpretations of measurement concept are analysed. One of those is purely quantum and is in  collection, for a given state of the object to be measured, of incompatible observable measurement results in amount enough for reconstruction of the state.  Two others make evident the difference between the reduced density matrix and the density matrices of physical objects involved in the measurement.

	It is shown that the von Neumann projectors, in combination with the concept of qudit phase space, produce an idea of a phase portrait of qudit state as a set of mathematical expectations for projectors on the possible pure states. The phase portrait is not a probability distribution since the projectors on non-orthogonal states are incompatible observables. Along with that, the phase portrait includes probability distributions for all the resolutions of identity of the qudit observable algebra.

	Additional peculiarities of measurement of the qudit degenerated observables, caused by the possibility of independent measurement of the observables for the particles, make possible to distinguish the local reduction of the qudit particle states from the entanglement of the local measurement results.
	
	The phase portrait of a composite system comprised by a qudit pair generates local and conditional phase portraits of particles. The entanglement is represented by the dependence of the shape of conditional phase portrait on the properties of the observable used in the measurement for the other particle.
	
	Analysis of the properties of a conditional phase portrait of a multiqubit qubits shows that absence of the entanglement is possible only in the case of substantial restrictions imposed on the method of multiqubit decomposition into qubits. Such a special method for determination of particles exists for each multiqubit state.

\end{abstract}

\section{Introduction}
The concept of qubit evolved \cite{Schumacher95}  as a result of attempts to expand the information theory to include the particles characterized by discrete set of states.
Logical structure of the process of information transfer \cite{shannon1} looks like that for the quantum measurement \cite{vonneumann}, and application of quantum carrier makes the measurement to be the main tool for getting information. 
In the real processes of information transfer and processing multiqubits, the quantum particles with $ d=2^p $ degrees of freedom, are usually present.

Measurement of qubits, multiqubits and other qudits differs from a typical measurement in quantum mechanics as much as the qudit properties differ from the properties of common particles of quantum mechanics. 
All the observables of a qudit are bounded and are represented by Hermitian matrices, unlike typical observables of quantum mechanics, such as coordinate and momentum, that are unbounded and representable by self-adjoint operators with continuous spectrum.
The main concepts of the qubit theory are induced by measurement reduction of state and absence of cloning; those replace the quantum-mechanical Schroedinger Cat paradox and the Heisenberg uncertainty principle \cite{nielsen, WheelerZurek}.

The process of multiqubit measurement is additionally complicated by the entanglement that is in dependence of probability of measurement for one part of the multiqubit on the device used for measuring the other part \cite{CERF1998}. 

The entanglement became topical with evolvement of quantum cryptography \cite{ekert91}, \cite{Bennett96A}. After successful development of entanglement theory for a qubit pair  \cite{PHPLA,horodecki:865}, more and more attention is paid to entanglement of several qubits  \cite{Mancini202010,Martin201812,Guhne201807}. In particular, possibility of quantitative evaluation of entanglement for several qubits \cite{Vedral201709,Adesso201608,Misra201608}, or, at least, possibility of detection for total entanglement \cite{Zukowski201112} are considered.   Especially efficient is the tomographic approach \cite{Mancini2008,Mancini2015} that makes it possible to characterize the qudit state by a set of probability distributions.

Here main concepts of multiqubit and qudit measurement theory are presented in form appropriate for practical application.
The report is organized as follows.

Section \ref{QO} deals with the properties of qudit measurement by non-generate observables. 
A new instrument for this measurement theory -- the phase portrait -- is introduced for practical application; this is a function on the phase space with the value of probability of registration, for a given qudit state, of pure state that corresponds to the phase space point under consideration.

Subsection \ref{QM} points out the physical differences between the act of measurement, a series of acts of measurement with same devices, and the measurement of quantum state. 

Subsection \ref{qN} deals with application of the concept of resolution of identity to the theory of qudit measurement. Here attention is paid to necessity of three different quantities for calculation of the expected value for a nondegenerate observable. The first one is the matrix of observable characterising the device, the second is the measured state density matrix, and the third is the matrix of the unitary transformation of resolution of identity for the observable to resolution of identity for the density matrix.

The last subsection \ref{qB} illustrates all the above by the example of a qubit. The phase portrait of a qubit is determined by a three-component vector, this is called here a director of the state phase portrait, hereinafter it is used in the analysis of the properties of conditional phase portrait of multiqubit particles.

Section \ref{DO} is related to the composite states of a qudit pair. First-priority attention is given to existence of nondemolition measurement a pair state. In the subsection \ref{DL} resolution of identity for the composite induced by the particle measuring devices and resolution of identity for a composite arbitrary state are constructed. Those resolutions of identity produce such particle observables that the nondemolition measurement for particles remains nondemolition for the composite as well. The subsubsection \ref{DF} deals with the conditional phase portraits of the particles; the properties of the last ones indicate that entanglement of measurement results is a common fact. 

Subsection \ref{DR} deals with separation of reduction of the results of composite qudit measurement into local reduction and entanglement. The last one is characterized by impossibility to obtain devices able to perform nondemolition measurement with local reconstruction of measuring devices.

Subsection \ref{DQ} includes demonstration of qudit pair theory by means of a well-described pair of qubits. In spite of pure states, for the mixed state of a qubit pair not only complete entanglement of results is specific, but a light entanglement of the results as well. 

In section \ref{DM} main aspects of the multiqubit measurement theory are considered. The algorithm for solving the problem on multiqubit state reconstruction by the results of $ 4^p -1$ series of measurements for qubit observables comprising a sufficient set of multiqubit incompatible observables is constructed. The properties of the conditional phase portrait and the properties of the measured state resolution of identity transformation to the observables resolution of identity matrices are used to analyse maximally entangled pure and mixed states.

\section{One Particle}
\label{QO}

At the beginning of this section the properties of the process of measurement that are common for all the particles with finite dimension of Hilbert space $ N $ are briefly characterized. The first property is that for each qudit state there exists a set of separable observables. 

\begin{itemize}
	\item 
	Mathematical substantiation of the statement about existence of separable observable is in the fact that an arbitrary Hermitian or unitary matrix can be represented by a weighted sum of projectors on eigenvectors.
	
	The Hermitian matrix represents an arbitrary observable $ A $ or a density matrix $ \rho $, and the unitary matrix $ U $  represents transformation in Hilbert space:
	\begin{equation}\label{MatrEigen}
		\begin{array}{ll}
			A=\sum\limits_{k=0}^{N-1}A_k\roa{k;A},&\\
			U=\sum\limits_{k=0}^{N-1}e^{i\phi_k}\roa{k;U},& \sum\limits_{k=0}^{N-1}\phi_k =0,\\
			\rho=\sum\limits_{k=0}^{N-1}p_k\roa{k;\rho},&\sum\limits_{k=0}^{N-1}p_k=1,\quad\forall_{ k \in \left[0,N-1 \right] }:  p_k\geq 0 .
		\end{array}
	\end{equation}
	
	The set of eigenvectors of a Hermitian  or a unitary matrix $ {X} $,
	\begin{equation}\label{Abasis}
		\mathfrak{b}_{X}=\left\lbrace \cat{k;{X}}, k \in \left[0,N-1 \right]\right\rbrace,
	\end{equation}
	forms the generating basis of the matrix, and the sequence of projectors on eigenvectors,
	\begin{equation}\label{MesPk}
		\mathfrak{R}_{X}=\left\lbrace\Pi_k\left[{X} \right] = \roa{k;{X}}, k \in \left[0,N-1 \right]\right\rbrace,
	\end{equation}
	is resolution of identity for the matrix $ {X} $. The projectors $ \Pi_k $ perform reduction of state by von Neumann \cite{vonneumann} in the process of measurement of the observable \eqref{MatrEigen}.
	
	\item 
	Physical substantiation of the statement about existence of separable observables is based on the fact that the projector $ \Pi $ is the mathematical representation of a counter. Really, in each act of measurement the device represented by the projector generates the value 1 or the value 0,  the average value $\overline{\Pi}=\nu_{\Pi}\equiv\frac{K_{1}}{K} $ in the measurement series of length  $ K $  is the frequency of successful events $ K_{1} $ and generates the evaluation of probability $ \lim_{K\mapsto \infty}\nu_{\Pi}= p_{\Pi} $ that is calculated as mathematical expectation of the projector  $p_{\Pi}=\tr{ \Pi\rho} $ for a specific state $ \rho $.

\end{itemize}

In this section only nondegenerate observables are considered, those are represented by informationally complete positive measures \cite{Weigert2006}, thus the resolution of identity  $\mathfrak{R}_{A}  $ consists of the projectors on pure states.  

Considerations given above make it possible to assert the existence of qudit observable with resolution of identity $\mathfrak{R}_{A}  $ same to the qudit density matrix resolution of identity $\mathfrak{R}_{\rho}  $. This observable provides the possibility of separating measurements for qudits.

\textbf{The qudit phase space} $ \mathfrak{P} $  consists of all the pure states of the qudit. It has the topology of the  sphere $ S_{2N-2} $  \cite{Glaser2020,Baumgartner_2007} and has the standard measure on the sphere $ d\mu_N $ normalized to identity matrix:
\begin{equation}\label{MesMes}
	\int\limits_{\mathfrak{P}}d\mu_N\left( \psi\right) =1\quad\mapsto\quad 
	\int\limits_{\mathfrak{P}}\roa{\psi}d\mu_N\left( \psi\right) =\hat{I}.
\end{equation}
Mathematical expectation of the projector to each pure state is the phase portrait of a state with density matrix $ \rho $:
\begin{equation}\label{MesPort}
	p\left[ \rho\right]\left(\psi \right)=\tr{\Pi_{\psi}\rho}\equiv \mess{\psi}{\rho},\quad \forall \psi \in \mathfrak{P}.
\end{equation}
The phase portrait includes all the probability distributions that can be obtained by measuring an observable, it is normalized to unity,
\[ \int\limits_{\mathfrak{P}}p\left[ \rho\right]\left(\psi \right)d\mu_N\left( \psi\right) =1, \]
though it is not a probability distribution since the counters $ \Pi\left(\psi \right)  $, $ \Pi\left(\varphi \right)  $  with nonorthogonal projectors $ \brcat{\psi}{\varphi}\neq 0 $ are not compatible observables.

The linear combination of the projectors of orthogonal resolution of identity \eqref{MesPk} forms an Abel subalgebra of compatible observables. Probability distribution on the resolution of identity and mathematical expectation $ \aver{A} $ for each of those observables in a given state with density matrix $ \rho $ are determined by the values $  p\left[ \rho\right]\left(\psi_k \right)$ of phase portrait on the resolution of identity:
\begin{equation}\label{MesPro}
	P_{X}\left[ \rho\right] =\left\lbrace p\left[ \rho\right]\left(\psi_k \right), k \in \left[0,N-1 \right]\right\rbrace,\quad 
	\aver{A}=\tr{A\rho}=\sum\limits_{k=0}^{N-1}A_kp\left[ \rho\right]\left(\psi_k \right)
	.
\end{equation}

The values of phase portrait on projectors $\Pi_k$ \eqref{MatrEigen} of the density matrix $ \rho $ resolution of identity  $ \mathfrak{R}_{\rho} $   are equal to eigenvalues of density matrix:
\begin{equation}\label{MesPortRho}
	p\left[ \rho\right]\left(k;\rho \right)=\aver{\Pi_k\rho}= p_k \quad \forall \roa{k;\rho} \in \mathfrak{R}_{\rho}.
\end{equation}
The measurement of an observable with resolution of identity $ \mathfrak{R}_{\rho} $ is a separating one.

\subsection{ Meanings of Measurement Concept }
\label{QM}
The second common for quantum measurements property is that the term  'measurement' has three different meanings.

\begin{itemize}
	\item 
	According to the first, physical meaning, the term 'measurement' represents the process of obtaining one of possible values $A_k $  of the observable that corresponds to the measuring device. This process consists of following stages. At the first stage the measured state of the particle with density matrix $ \rho\sgn{P} $  is prepared, and the state $ \rho\sgn{D}_{in} $ of the device is adjusted. Then the stage of interaction of the particle and the device takes place as a time-dependent unitary transformation of composite density matrix of the system  'particle + device':
	\[ \rho\sgn{P\&D}\left( t\right) =U\left(t \right) \rho\sgn{P}\times\rho\sgn{D}_{in}U^+\left(t \right). \]
	At the next stage registration of results takes place. The composite density matrix is just split into product of one of the projectors $ \Pi_{count}\left[A \right] $ of resolution of identity for the observable and the corresponding density matrix of the device $ \rho\sgn{D}_{count} $:
	\[ \rho\sgn{P\&D}\left( t_{count}\right) = \Pi_{count}\left[A \right]\times\rho\sgn{D}_{count}. \]
	Registration not only determines the state of the device $ \rho\sgn{D}_{count} $ along with the number of the projector $ \Pi_{count}\left[A \right] $ that determines the following behaviour of the measured particle. It finalizes the physical process of measurement for a given sample of measured particle. The following act of measurement deals with another sample of the particle.

	Thus, a separate act of measurement leads to a dual result, with the number $ {count} $ of one of eigenstates of the observable and the particle in registered state with number $ {count} $.  The following behaviour of the particle does not affect the result, irrespectively of either the particle and the device recur to the initial states according to the prediction of the Poincar\UseTextAccent{OT1}{\'}{e} theorem, or not.

	\item 
	The second, statistical meaning of the term 'measurement' represents the process of obtaining results for a series of independent acts of measurement. Independence means that the qudit states at the process of registration in different measurement acts cannot form superpositions and in general there is no physical object common for the whole series. It is so, for instance, because of realization of measuring acts with one device at different time instants or with different spatially separated similar devices. The result of the series of measurements forms a sequence of resolution of identity projector numbers for a measured observable.  Part of information present in this sequence can be represented by the density matrix averaged by the whole sequence. If $ \nu_k=\frac{P_k}{P} $ is the frequency of repetition of the value $ k$ in the sequence of results, the averaged density matrix is
	\[ \bar{\rho}=\sum_{k=1}^{K}\nu_k\Pi_k\left[A \right]. \]
	The reduced density matrix is the value
	\begin{equation}\label{Qreduc}
		\rho\sgn{rd}=\sum_{k=0}^{N-1}p_k\left[A \right]\Pi_k\left[A \right], \quad 
		p_k=\tr{\rho\Pi_k\left[A \right]}.
	\end{equation}
	This is mathematical expectation of the density matrix averaged by the series of measurements for the qudit in state $ \rho $  observable $  A$.	There are no reasons to consider   $ \rho\sgn{rd} $ as the state density matrix for the measured particle, with exception of a pure state and the measurement being non-destructive; in such a case the reduced density matrix is equal to the density matrix of measured state.
	\item 
	The third meaning is specific to quantum measurements only. Some problems, first of all the problem on reconstruction of state, require measurement of statistical properties of more than one incompatible observable. In such a case reduced density matrices are different for different observables.  The measurement of a set of incompatible observables differing by a continuous parameter is called 'method of quantum tomography'  \cite{Vogel1989,Raymer1993,Mancini1996}.
\end{itemize}

A set of measurements provides other possibilities as well. For instance, it is easy enough to select all the terms of the sequence for which the state number is equal to a given one. If after measuring the qudit is involved to other processes, the measuring device is a generator of a given state of qudit for those following processes.

The mixed state of qudit can be a natural one if the qudit is a part of a more complicated system. In this case measurement of the other part of such a complicated system can affect the results of individual measurement acts in different ways. The other method for measuring mixed states is in preparation of pure states with repetition frequency approximating required probabilities. Shortcoming of this method is in possibility to control which state is prepared for a specific measurement act, and  how to exclude randomness.

\subsection{Qudit Description}
\label{qN}

\label{spaces}
One more common peculiarity of qudit measurements is in involvement of quantities forming several different spaces.

The qudit pure states are more often represented by normalized vectors of complex Hilbert space  $\mathcal{H}_N=C^N$. Dimension of this space is determined by the qudit dimension. One more space is the qudit phase space that consists of all the pure states, this is topologically equivalent to the sphere $\mathcal{P}_N= S_{2N-2} $.
Dimensions of the most of other spaces associated to qudit are proportional to $ N^2 $.

Algebra of observables is a vector space with topology $R^{N^2}$; the special unitary group  $SU(N)$ is a compact topological space locally equivalent to $R^{N^2-1}$; the space of density matrices is a compact area in $R^{N^2-1}$, restricted by inequalities for the eigenvalues; the resolution of identity space $ \mathcal{B}_N $ has dimension $ N^2-N $ with topology of direct product of $ N $ spheres with dimensions that uniformly decrease from $ 2N-2 $ to 2. An exception is the space of the density matrix eigenvalues, this is a simplex $ T_{N-1} $ with topology of a compact  simply connected region in $R^{N-1}$.

\subsubsection{Qudit Spaces}

The set $ \mathcal{B}_N $ of the Hilbert space $\mathcal{H}_N$ orthonormal bases with typical element $ \mathfrak{b}  $,
\begin{equation}\label{Abas}
	\mathfrak{b} = \left\lbrace \cat{k},\ k \in\left[0,N-1 \right],\quad \forall m\in\left[0,N-1 \right]\  \brcat{k}{m}=\delta_{k,m}  \right\rbrace,
\end{equation} 
is equivalent to the set of qudit resolution of identity $ \Pi\left[\mathfrak{b} \right]  $ with typical element $\Pi\left[\mathfrak{b} \right]  $,
\begin{equation}\label{Aproj}
	\Pi\left[\mathfrak{b} \right] = \left\lbrace \Pi_k=\roa{k},\ k \in\left[0,N-1 \right],\quad  \forall m\in\left[0,N-1 \right]\   \Pi_k\Pi_m=\Pi_k\delta_{k,m}  \right\rbrace.
\end{equation} 
A projector of a resolution of identity $ \Pi_k $ is an element of a phase space with topology of sphere $ S_{2N-2} $, all the other projectors of this resolution of identity are orthogonal to this projector  $ \Pi $ and form a resolution of identity for a qudit with dimension $ N-1 $, thus the space $ \mathcal{B}_N $ has a real dimension $ N^2-N $ and is topologically equivalent to the direct product of a sequence of spheres: 
\begin{equation}\label{SpaceBasis}
	\mathcal{B}_N\equiv S_{2N-2}\times S_{2N-4} \times \ldots \times S_2 .
\end{equation}

The space $ \mathcal{B}_N $, like the phase space $\mathcal{P}_N$, is homogeneous with respect to the group $ SU(N) $, thus there exists a pair of conjugate matrices
\begin{equation}\label{AtoRho}
	U\left[A\mapsto\rho \right] =\sum_{k=0}^{N-1}\ro{k;\rho}{k;A},\quad U\left[\rho\mapsto A\right]=U^+\left[A\mapsto\rho \right]
\end{equation}
transforming the basis  $ \mathfrak{b}_A $ and the resolution of identity $\Pi\left[\mathfrak{b}_A \right]  $ of the observable $A  $ to basis $ \mathfrak{b}_{\rho} $ and resolution of identity $\Pi\left[\mathfrak{b}_{\rho} \right]  $ of density matrix $ \rho $ and vice versa:
\[ \Pi_k\left[\rho \right]=U\left[A\mapsto\rho\right]\Pi_k\left[A \right] U^+\left[A\mapsto \rho\right],\quad \Pi_k\left[A \right]=U\left[\rho\mapsto A\right]\Pi_k\left[\rho \right] U^+\left[\rho\mapsto A\right]. \]

The matrix elements of the matrices \eqref{AtoRho}
\[ U\left[A\mapsto\rho \right]_{k,k'} =\brcat{k;A}{k';\rho},\quad
U\left[\rho\mapsto A\right]_{k,k'}=\brcat{k;\rho}{k';A}
\]
determine, together with the eigenvalues of the observable $ A_k $ and of the density matrix $ p_{k} $, the expected value $\aver{A}=\tr{A\rho}  $ of the observable $ A $ in state $ \rho $:
\begin{equation}\label{MathARho}
	\aver{A}=\tr{A\rho}=\sum_{k=0}^{N-1}A_k\mess{k;A}{\rho}=\sum_{k,k'=0}^{N-1}A_kp_{k'}\left|U\left[A\mapsto\rho \right]_{k,k'}\right|^2 .
\end{equation}
Thus, the value of observable in a given state is determined by three quantities: the eigenvalues of the observable $ \left\lbrace A_k,\ k \in\left[0,N-1 \right] \right\rbrace  $, the eigenvalues of the density matrix $ \left\lbrace p_k,\ k \in\left[0,N-1 \right] \right\rbrace  $ and the matrix \eqref{AtoRho} of the mutual transformation of the resolutions of identity for the density matrix and for the observable.

The boundaries of the density matrices space  $ \mathcal{R}_N$ are determined by the conditions of  eigenvalues nonnegativity. The eigenvalues form a simplex $ T_{N-1} = \left\lbrace p_k\geq 0,\ k \in \left[0,N-1 \right], \ \sum p_k=1  \right\rbrace $; the projectors $ \left\lbrace \roa{k;\rho},\ k \in \left[0,N-1 \right]   \right\rbrace  $ on eigenvectors are the elements of the resolution of identity space $ \mathcal{B}_N $. Thus, the space $ \mathcal{R}_N=T_{N-1} \times \mathcal{B}_N $ of density matrices is the direct product of the simplex $ T_{N-1} $ of density matrix eigenvalues and the space of resolutions of identity  $ \mathcal{B}_N $. 
The simplex of eigenvalues is restricted by the facets $ p_k=0 $, the set of unitary transformations for an arbitrary facet forms the boundary of the space $ \mathcal{R}_N$ of density matrices. The vertices of the simplex of eigenvalues correspond to pure states. The set of vertices is equivalent to the qudit phase space $\mathcal{P}_N= S_{2N-2} $.
\subsubsection{Extension of Pauli Matrices}

Algebra of Hermitian matrices is a real vector space with dimension $ N^2 $ and with scalar product $ A\cdot B =\tr{AB} $. 

The induced basis  $ \mathfrak{b}\sgn{Ind} =\left\lbrace \ro{m}{n},\  n,m\in \left[0,N-1 \right]\right\rbrace  $ of the matrix space over the Hilbert space $H=C^N$ is produced by each basis $ \mathfrak{b}=\left\lbrace \cat{n},\  n\in \left[0,N-1 \right]\right\rbrace  $ of the qudit Hilbert space.

Non-diagonal matrices of the induced basis,
\begin{equation}\label{Ladder}
	a\sgn{n,m}=\ro{n}{m},\quad {a^+}\sgn{n,m}=\ro{m}{n},
\end{equation}
are often called the ladder  operators or the creation/annihilation  operators,  since those are used to describe the processes of excitation transfer from one degree of freedom $ \cat{m} $ to another one $ \cat{n} $. The adjoint matrices $ {a^+}\sgn{n,m} $ are responsible for the reverse transfer.

The space of Hermitian matrices is the subspace of the algebra of complex matrices $N\times N$  and has a two times smaller dimension, thus application of the induced basis is ineffective as to the number of parameters. One of the bases of the set of Hermitian matrices as a vector space is formed by the Pauli matrices associated to each pair of degrees of freedom $ n $ and $m $:
\begin{equation}\label{PauliNM}
	\begin{array}{l}
		\forall\ 0\leq m < n\leq N-1: \\
		\sigma\sgn{n,m}_x=\ro{n}{m}+\ro{m}{n},\quad \sigma\sgn{n,m}_y=i\ro{n}{m}-i\ro{m}{n},\\
		\sigma\sgn{n,m}_z=i\sigma\sgn{n,m}_x\sigma\sgn{n,m}_y\equiv \Pi_{n}-\Pi_{m}.
	\end{array}
\end{equation}
Multiplication rules
\begin{equation}\label{PauliNMprop}
	\sigma\sgn{n,m}_a\sigma\sgn{n,m}_b=\left(\Pi_{n}+\Pi_{m} \right) \delta_{ab}+i\epsilon_{abc}\sigma\sgn{n,m}_c
\end{equation}
complete the matrices \eqref{PauliNM} up to the basis of the algebra of observables.

The sequence $ \left\lbrace b_p, p\in \left[ 0,N^2-1\right]  \right\rbrace  $ of $ N $ diagonal matrices $ \Pi_{n} $ and $ N^2-N $ non-diagonal Pauli matrices
\begin{equation}\label{Gbasis}
	b_p=\left\lbrace \begin{array}{ll}
		\sigma\sgn{n,m}_x:& n<m\\
		\Pi_{n}=\roa{n}:&n=m\\
		\sigma\sgn{n,m}_y:& n>m\\
	\end{array}\right|  ;\quad p=n+m\cdot N,\ n,m \in \left[0,N-1 \right] 
\end{equation}
forms the basis of the space of observables as a real space $R^{N^2}$. 
This basis differs from the one described in \cite{Munro2002} by choice of diagonal matrices only.

More convenient in following application for qudit clustering is a basis in which one of diagonal elements is the identity matrix $\beta_0=\hat{I}  $, the non-diagonal ones are Pauli matrices \eqref{PauliNM}; the other diagonal elements $\beta_{m\left(N+1 \right)}  $ are the traceless linear combinations of matrices $ \Pi_{n} $:
\begin{equation}\label{Basis}
	\begin{array}{l}
		\beta_{m\left(N+1 \right)}=\sum\limits_{n=0}^{N-1}C^{-1}_{m,n}\Pi_{n},\\
		\Pi_{n}=\frac{1}{N}\beta_0+\sum\limits_{m=1}^{N-1}C_{m,n}\beta_{m\left(N+1 \right)}
	\end{array}
	:\quad \tr{\beta_p}=0,\quad
	\tr{\beta_p\beta_q}=M_p\delta_{p,q}.
\end{equation}
In this basis an arbitrary observable $  A $ and an arbitrary density matrix $ \rho $ are:
\begin{equation}\label{ObsInBasis}
	A=A_0\hat{I}+\sum_{p=1}^{N^2-1} A_p  \beta_p,\quad\rho=\frac{1}{N}\hat{I}+\sum_{n=1}^{N^2-1} \frac{d_p }{M_p} \beta_p.
\end{equation}
The coefficients $ d_p $ characterize the deviation of density matrix from the equilibrium one $ \frac{1}{N}\hat{I} $ and determine the expected values of observables represented by matrices $ \beta_p $,
\begin{equation}\label{BasisCoeff} 
	\aver{\beta_p }=\tr{\beta_p\rho} = d_p.
\end{equation}
In the basis of eigenvectors of the observable only  $ A_p $ with indices $ p=m\left(N+1 \right)  $ corresponding to diagonal matrices \eqref{Gbasis} and the linear combinations of those \eqref{Basis} are nonzero. Similarly, in the basis of eigenvectors of the density matrix only $ d_p $ with the same indices are nonzero.

The average value of the observable $ A $ in the state with density matrix $ \rho $ is a scalar product of the vector of the observable with components $A_p$ and the vector of the density matrix with components $ d_p $:
\begin{equation}\label{BasisAver}
	\aver{A}\left[ \rho\right] =A_0+\sum_{p=1}^{N^2-1} A_p  d_p.
\end{equation}
Thus, to determine the density matrix  $ N^2-1 $ independent numbers are needed. Jointly are to be measured $ N-1 $ independent orthogonal projectors \eqref{MesPk}, therefore to determine the density matrix of an arbitrary state not less than $ N+1 $ incompatible resolutions of identity are needed. This statement is a somewhat more strict form of the no-cloning theorem \cite{Wootters82}.

\subsubsection{Qudit Symmetry}

The space of pure states is homogeneous with respect to the special unitary group  $ SU(N) $;  for each pair of pure states there exists a unitary matrix that transforms one state into another one. Dimension $ N^2-1 $ of this group as a topological space is noticeably larger than the dimension $ 2N-2 $ of the phase space, thus each state with vector $ \cat{\psi} $ has a corresponding stabilizer subgroup with dimension $ \left( N-1\right)^2  $, this is a group $ U(N-1) $ of unitary transformations for the subspace of vectors orthogonal to $ \cat{\psi} $.

The space of resolutions of identity $ \mathcal{B}_N $ is homogeneous with respect to $ SU(N) $ as well. The stabilizer subgroup of an arbitrary resolution of identity \eqref{MesPk} consists of unitary matrices
$ U=\sum e^{i\phi_k}\Pi_k$ ( $ \sum \phi_k=0 $) that commute with each projector $\Pi_k  $.

The generator $ J  $ of an arbitrary element $ U=\exp\left(i J \right) $ of this group is a Hermitian matrix with zero trace. Each generator produces an one-parameter Abel subgroup with elements
\begin{equation}\label{Utau}
	U\left(\tau \right) \left[J \right] =\exp\left(i\tau J \right).
\end{equation}
The subgroups with one of basis matrices \eqref{Gbasis} as a generator are periodical ones, since the exponent in  \eqref{Utau} for such subgroups takes trigonometric form:
\begin{equation}\label{UtauTrig}
	U\left(\tau \right) \left[\beta_p \right] =\hat{I}-\Pi\sgn{\beta}_p+\cos\tau  \Pi\sgn{\beta}_p+i\sin\tau \beta_p,\quad \Pi\sgn{\beta}_p=\frac{\beta_p^2}{M_p}.
\end{equation}

Real transformation of the basis vectors is performed by  $ N^2-N $ subgroups produced by generators $ \sigma\sgn{m,n}_{x,y} $ \eqref{PauliNM}. The orbit of each such subgroup in the space of resolution of identity $ \mathcal{B}_N $ is one of two perpendicular meridians of the Bloch sphere associated with the degrees of freedom $m $ and  $n $.

The generator and each element of the subgroup produced by it have a common resolution of identity,
\begin{equation}\label{GenJ}
	\begin{array}{c}
		J =\sum\limits_{n=0}^{N-1}J_n\roa{n;J},\\
		U\left(\tau \right) \left[J \right]=\sum\limits_{n=0}^{N-1}\exp\left(i\tau J_n \right)\roa{n;J},
	\end{array}
	\quad \sum_{n=0}^{N-1}J_n=0.
\end{equation}
Each term $ \exp\left(i\tau J_n \right)\roa{n;J} $ corresponds to rotation by $ \tau J_n$ with respect to the direction $ \cat{n;J} $.

Common generator produces a non-periodic subgroup (an exception is the generator with rational ratio of all the eigenvalues $ J_n/J_m $). Corresponding trajectories in the space of resolution of identity are unlimited, though in some cases those trajectories are fixed points of the subgroup or belong to its invariant subspace.

Dependence of the closed qudit state on time $ t $ is determined by the unitary matrix with generator linearly depending on time, $J=Ht$. Common Hamiltonian has no rational ratios of eigenvalues, thus the common state is aperiodic. Qubit is an exception, the qubit period is determined by the difference of two Hamiltonian eigenvalues. 

The vector representation \eqref{ObsInBasis} of the observables and the density matrices produces representation  of the special unitary group $ SU(N) $ with rotations of vectors $ A_n $ and  $ d_n $ from the group  $ O_{N^2-1} $ of rotations $R^{N^2-1}$. Dimension $ N^2-1 $ of the group  $ SU(N) $ is less than the dimension $ \left(N^2-1 \right) \left(N^2-2 \right)/2 $  of the group $ O_{N^2-1} $, thus the unitary transformations are not enough to direct the vectors of the basis along a given vector, though are enough for transformation of an arbitrary density matrix to a diagonal matrix. This transformation turns to zero all the components  $ A_n $ and  $ d_n $ of the respective vector, except of the components of diagonal basis matrices with indices  $ n=n'\left(N+1 \right)  $.

\subsubsection{Measures of Reduction}
Reduction of the measured state is present in each case when the matrix \eqref{AtoRho} of the unitary transformation $ U\left[A \mapsto\rho \right] $ of the device resolution of identity $ \mathfrak{R}_A $ to resolution of identity  $ \mathfrak{R}_{\rho} $ of the measured state density matrix does not belong to stabilizer subgroup.

Spectral representation \eqref{MatrEigen}  of the transformation matrix represents this matrix by a composition of rotations with respect to the eigenvectors $ \cat{n;U\left[A \mapsto\rho \right] } $ by angles $ \varphi_n $. Mathematically valid measure of reduction is the transformation matrix generator norm 
\begin{equation}\label{MesRed}
	M\sgn{Rd}	= \sqrt{\sum_{n=0}^{N-1}\varphi^2_n }.
\end{equation}
Effect of different angles depends on the measured state, since each angle characterizes the reduction of part of the states only. For instance, transformation by the matrix $U=\exp{\left(i\phi \sigma_x\sgn{01} \right) }$ entangles the measurement of states from the subspace formed by vectors $\cat{0}$ and $\cat{1}$, but the states from the subspace with basis $ \left\lbrace \cat{n},\ n\in\left[2,N-1 \right]   \right\rbrace $ are measured without changes. Thus, the number of nonzero eigenvalues of transformation matrix generator is an additional characteristic of the reduction of state.

Physically meaningful characteristic of the reduction of pure state is the entropy of the measurement result
\begin{equation}\label{MesS}
	S\sgn{mes}	= -\sum_{n=0}^{N-1}\nu_n\log_2 \nu_n
\end{equation}
that is nonzero only in the case of reduction. Reduction of measurement of a mixed state leads to increase of entropy of the measurement by a value substantially depending on the entropy of the state.

\subsubsection{Phase portrait}

The phase portrait of the qudit is easily calculated in the basis of the density matrix eigenvectors.

\textbf{Phase portrait of a pure state} is determined by a scalar product of the state vector $ \cat{\phi} $ and the vector $ \cat{\psi} $ on which the measured projector $ \Pi\left[\psi \right] $ projects:
\begin{equation}\label{PortraitPure}
	p\left[\phi \right]\left( \psi\right) =\tr{\roa{\phi}\Pi\left[\psi \right] }=\left|\brcat{\phi}{\psi} \right|^2=\cos^2\theta/2=\frac{1+\cos\theta}{2}.
\end{equation}
Here $ \theta $ is the distance in phase space between the points representing the state and the measured projector.

\textbf{Phase portrait of a mixed state} is a weighted sum of phase portraits of generating states:
\begin{equation}\label{PortraitMix}
	p\left[\rho \right]\left( \psi\right) =\sum_{k=0}^{N-1}p_k\tr{\roa{\phi_k}\Pi\left[\psi \right] }=\frac{1}{2}+\sum_{k=0}^{N-1}\frac{p_k\cos\theta_k}{2}.
\end{equation}
Since the eigenvectors of the density matrix are orthogonal, the direction cosines to all the other vectors turn to zero when the director of the measured projector approaches the director of one of eigenstates. As the result, at each eigenvector of density matrix there is one local maximum $ \frac{1+p_k}{2} $ and one local minimum $ \frac{1-p_k}{2} $ of the phase portrait.

\subsection{Qubit }
\label{qB}

Quantum measurement of qubit, as the simplest quantum object, has almost all peculiarities of the quantum measurement of qudit, except of ability of division into particles.

\subsubsection{Qubit spaces}
Most elements of the basic space $H=C^2$ are redundant, the qubit pure states are represented only with normalized vectors $\cat{\psi}$, the set of which is equivalent to a two-dimensional sphere $\mathfrak{P}_2=S_2$. This sphere, it is the Bloch sphere, is the qubit phase space, its points are parametrized by a pair of spherical coordinates $ \left\lbrace\vartheta,\varphi \right\rbrace  $ or by a three-dimensional unit vector $ \vec{n}=\left\lbrace \sin\vartheta\cos\varphi,\sin\vartheta\sin\varphi,\cos\vartheta \right\rbrace  $.

The space of observables $\mathfrak{O}_2$  consists of Hermitian $ 2\times2 $ matrices. The Pauli matrices \eqref{PauliNM}, together with the identity matrix, form the basis of the algebra of observables as a real vector space: 
\begin{equation}\label{QA}
	A =A_0\hat{I}+\vec{A}\cdot\vec{\sigma}.
\end{equation}
The parameters $A_0$, $\vec{A}=\left\lbrace A_x,A_y,A_z \right\rbrace $ are arbitrary real numbers, the algebra of observables $\mathfrak{O}_2=R^4$ is a four-dimensional real vector space. Application of vector notations for the elements of the three-dimensional vector space is related to standard notations in elementary physics. Real spatial meaning those vectors have in the case of qubits characterizing the spin states. In all the other implementations of the qubit spatial interpretation is absent.

The vector $ \vec{A} $ is represented by the product of the norm $ A=\sqrt{\vec{A}^2} $ and the unit vector $ \vec{m}=\frac{\vec{A}}{A} $. The director $ \vec{m} $ of the observable determines the direction to the points of the Bloch sphere corresponding to the eigenvectors of the observable. Resolution of identity for the observable is formed by the pair of projectors 
\begin{equation}\label{QAnorm}
	\Pi_0\left[ A\right]=\frac{1}{2}+ \frac{1}{2}\vec{m}\cdot\vec{\sigma},\quad
	\Pi_1\left[ A\right]=\frac{1}{2}- \frac{1}{2}\vec{m}\cdot\vec{\sigma}.
\end{equation}
A qubit observable differs from a counter by the scale multiplier $ 2A $ and the shift of the reading by the value $ A_0 +A $.

The elements of the space $\mathfrak{B}_2$ of orthogonal resolutions of identity are the orthogonal pairs of projectors \eqref{QAnorm}. Each pair is enough determined if the normalized eigenvector $ \cat{\vec{m}} $ of one of projectors determined by the unit vector $ \vec{m} $ , the projector director, is known.

One more space $\mathfrak{R}_2$ is formed by mixed states. Those can either be formed artificially, as a combination of a set of pure states into a statistical ensemble, or can come into existence naturally if the qubit is a part of a more complicated object.

An arbitrary mixed state  is represented by a density matrix that can be written as a linear combination 
\begin{equation}\label{Qstate}
	\rho\left( {\vec{d}}\right) =\frac{1}{2}\hat{I}+\frac{1}{2}\vec{d}\cdot\vec{\sigma}
\end{equation}
of three Pauli matrices.

The vector $ \vec{d}=d\vec{n} $, $ \left| \vec{n}\right|=1  $, $ d\leq 1 $ determines the directions to the points on the Bloch sphere that correspond to the eigenstates of the density matrix
\[ \Pi\left( {\vec{d}}\right)=\frac{1}{2}\hat{I}+\frac{1}{2}\vec{n}\cdot\vec{\sigma},\  \Pi\cat{0}=\cat{0},\quad
\Pi_1\left( {\vec{d}}\right)=\frac{1}{2}\hat{I}-\frac{1}{2}\vec{n}\cdot\vec{\sigma},\  \Pi_1\cat{1}=\cat{1}.
\] 
The eigenvalues of the density matrix are $p=\frac{1+d}{2} $,  $p_1=\frac{1-d}{2}=1-p $, thus the matrix can be represented by the sum
\begin{equation}\label{QRho}
	\rho\left( {\vec{d}}\right)=p \Pi\left( {\vec{d}}\right)+p_1 \Pi_1\left( {\vec{d}}\right)=\left(1-p \right)\hat{I}+\left(2p-1 \right)\Pi\left( {\vec{d}}\right)  .
\end{equation}

The vicinity of the point $\vec{d}=0  $ is similar to the vicinity of the origin of a vector space $R^3$.  The space of the density matrices $\mathfrak{R}_2$ is a unit ball $ d\leq 1 $ with boundary $ d= 1 $ that is the Bloch sphere of pure states. The second form in the representation of the density matrix \eqref{QRho} shows that the qubit density matrix is determined by one projector. 

One more space associated to the qubit is a special unitary group $ SU(2) $ that keeps unchanged the mapping of state vectors into phase space. An arbitrary matrix of this group is represented by exponent 
\begin{equation}\label{Uexp}
	U=e^{iJ},
\end{equation}
with generator  $ J $ that is a traceless Hermitian matrix, thus it is represented by a combination of Pauli matrices 
\begin{equation}\label{UexpJ}
	J=\vec{j}\cdot\vec{\sigma}=j\vec{n}_j\cdot\vec{\sigma}.
\end{equation}
Here $ j$ is length, and the unit vector $\vec{n}_j $ -- is the direction of the generator vector.

Square of generator is proportional to the unit matrix $J^2=j^2\hat{I}$, so the exponential representation \eqref{Uexp} becomes a trigonometric one,
\begin{equation}\label{Utrig}
	U=e^{iJ}=\cos j\ \hat{I}+i\sin j\ \vec{n}_j\cdot\vec{\sigma}.
\end{equation}

\subsubsection{Phase portrait}

The qubit density matrix \eqref{Qstate} with director  $ \vec{d}=\left(p-1/2 \right)\vec{n}  $ determines the phase portrait of the qubit
\begin{equation}\label{PortraitQ}
	p\left[ \rho\right]\left(\theta \right)  = \frac{1}{2}+\left(p-\frac{1}{2} \right)\vec{n}\cdot\vec{m}  = \frac{1}{2}+\left(p-\frac{1}{2} \right)\cos\theta.
\end{equation}
Here $ \vec{m} $ determines the point of phase space, and $ \theta $ is the angle between the director of state and the director of the projector.

The phase portrait of the qubit has an axis of symmetry. This is the diameter of the Bloch sphere connecting the points that represent the eigenvectors of the density matrix. It is directed along the direction of the state director $ \vec{n} $, thus this vector can be called a director of phase portrait.

\begin{figure}[h]
	\centering
	\includegraphics[width=0.3\linewidth]{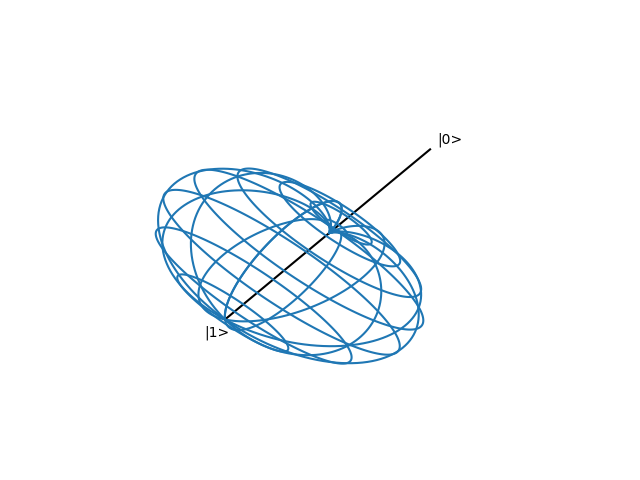}\includegraphics[width=0.3\linewidth]{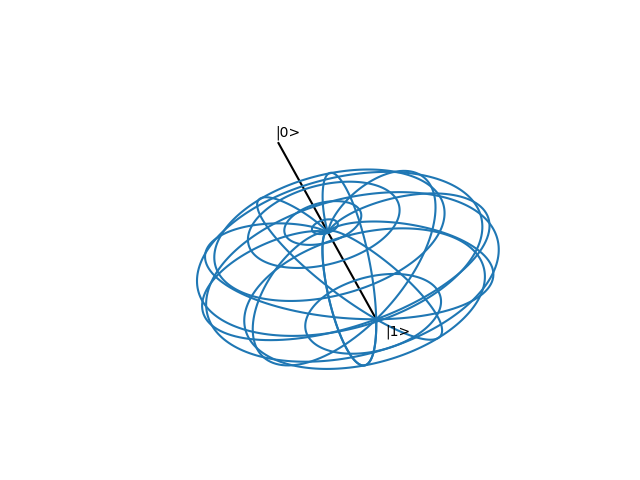}\includegraphics[width=0.3\linewidth]{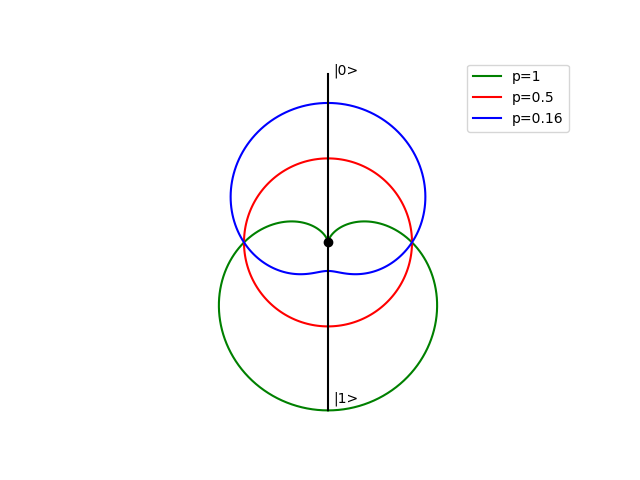}
	\vspace*{8pt}
	\caption[Qubit Predict]{Examples of qubit phase portraits. Left and centre -- phase portraits of qubit pure states with two different positions of state vectors on Bloch sphere. Probability is one in the case of detector,  direction of which coincides with the direction of the density matrix vector (the lower pole of the phase portrait), and zero in the case of detector with opposite direction (the upper pole of the phase portrait).
		Right -- cross-sections of qubit phase portraits for three listed values of state weight.
	}
	\label{fig:phpqubit}
\end{figure}

The shape of the generatrix  of the surface of phase portrait depends on the part $p$ of the state $ \cat{1;\rho}  $ in the mix. The right graph in figure \ref{fig:phpqubit} shows several generatrices for different values of probability. All the generatrices intersect in the point that belongs to diameter normal to the director of qubit state and is at distance 1/2 from the centre. Thus, the result of measurement for the observable directed on the Bloch sphere normally to the state director has probability 1/2 irrespectively of probability distribution for registration of eigenstates specific for this state.  The circle corresponds to equilibrium state; the phase portrait of pure state has the lower pole at distance 1 from the centre, the upper one is in the centre; the phase portrait of the mixed state has the lower pole at distance $ p $, this is equal to the state $ \cat{1;\rho}  $ score, from the centre, and the upper pole at distance $ 1-p $ in the direction of state $ \cat{0;\rho}  $.

\subsubsection{Qubit Measurement}
Specific character of measurement for qubit observables results from the fact that an arbitrary resolution of identity for qubit is determined by one projector, thus each observable is a linear function of a counter $\Pi$ characterised by the director $\vec{m} $ as $ \Pi_0\left[ A\right]$ in \eqref{QAnorm}.

Each separate act of measurement results in somewhat equivalent to No or Yes, and this gives one bit of information. 

A series of measurement acts for one observable produces repetition frequency $ \nu=K_1/K $ of successful readings for the counter, this is evaluation of probability $ p=\tr{\Pi\rho} $ and is given by the value  \eqref{PortraitQ} of the measured state $ \rho $ phase portrait in the point corresponding to the counter $  \Pi$.

Reduction of state at measurement is determined by the angle $ \theta $ between the vectors of state $ \vec{n} $ and the counter $ \vec{m} $ directors.  Entropy of the measurement result
\begin{equation}\label{Qentr}
	S\sgn{Rd}=H\left[\frac{1}{2}+\left(p-\frac{1}{2} \right)\cos\theta\right] \geq S\sgn{N}=H\left[p\right]  
\end{equation}
is larger than the entropy of the state unless $ \theta $ is zero.

The problem of the qubit state reconstruction requires measurement of repetition frequencies for 3 incompatible counters. As the result, evaluation of three components of the vector $ \vec{d} $ of the state deflection from equilibrium one is obtained.

\section{Two Particles}

\label{DO}
The qudits of a qudit pair have Hilbert space dimensions  $N_L$ and $N_S$ and are characterized by phase spaces  $\mathcal{P}_{N_L}=S_{2N_L-2}$ and  $\mathcal{P}_{N_S}=S_{2N_S-2}$, simplexes of density matrices eigenvalues  $T_{N_L-1}$  and  $T_{N_S-1}$, spaces \eqref{SpaceBasis} of resolution of identity $ \mathcal{B}_{N_L} $ and $ \mathcal{B}_{N_S} $, groups of symmetry $SU(N_L)$ and $SU(N_S)$ and phase portraits $ P\sgn{L}:\ \mathcal{P}_{N_L}\mapsto \left[ 0,1\right]   $ and  $ P\sgn{S}:\ \mathcal{P}_{N_S}\mapsto \left[ 0,1\right]   $. The possible results of measurement for a pair of observables are predicted by a common phase portrait $ P\sgn{L\& S}:\ \mathcal{P}_{N_L}\times\mathcal{P}_{N_S} \mapsto \left[ 0,1\right]   $. This phase portrait is produced by a common observable $ \Pi\left[\phi\sgn{L} \right] \times\Pi\left[\phi\sgn{S} \right]  $ that is a direct product of the projectors on pure states  $ \cat{\phi\sgn{L}}$ and  $\cat{\phi\sgn{S}} $.

The composite qudit Hilbert space is formed by the Hilbert spaces of the two qudits, $ H\sgn{L\& S}=H\sgn{L}\times H\sgn{S} $, while its space of resolution of identity \eqref{SpaceBasis} $ \mathcal{B}_{N_LN_S} $ has the dimension larger by $N_LN_S\left(N_L+N_S \right) $ than the dimension of the product of spaces, $ \mathcal{B}_{N_L} \times \mathcal{B}_{N_S} $, and the group of symmetry  $SU(N_LN_S)$ has the dimension larger by $N_L^2+N_S^2  $ than the dimension of the product of groups,  $SU(N_L)\times SU(N_S)$. Evaluation of dimensions indicates that some properties of composite qudit cannot be represented through the properties of the particles.

According to \eqref{MathARho} the measurement of an arbitrary observable of a composite qudit is determined through the measured state density matrix eigenvalues $p_k$, the measured observable eigenvalues $A_k$ and the transformation matrix $ U\left[\rho\mapsto A \right]  $ of the density matrix resolution of identity $\mathfrak{R}_{\rho}  $ to the resolution of identity of the observable $\mathfrak{R}_{A}  $. Existence of the composite qudit group elements that are not products of the elements of local groups results in existence of measurements for which it is not possible to separate local results. Measurement of a composite qudit arbitrary state with the observables of particles can be:

\begin{itemize}
	\item 
	separating, in the case of locally separating observables and separation into particles matched with the state of the composite qudit;
	\item locally reducing, if separation into particles is matched with the state of the composite qudit;
	\item entangling.
\end{itemize}

Thus entanglement is a consequence of disbalance in resolutions of identity of measurement devices and the measured state.

\subsection{Local observables and local states}
\label{DL}
Combination of a qudit pair with formation of a composite qudit takes place through formation of joint Hilbert space with corresponding space of matrices, with following separation of composite qudit required subspaces -- the phase space, the space of resolution of identity, the space of density matrices, etc.

The phase space of a composite qudit includes as its subspace a composition of phase spaces of particles, thus the qudit pair phase portrait is a part of the composite qudit phase portrait.

The observables represent the devices that can be constructed and fabricated according to the purpose of measurement, while the states represent the properties that the measured object has, or would have.

\subsubsection{Composition of observables of the particles}

\label{DN}

Measurement of two qudits is as well a measurement of a composite qudit with Hilbert space  $ H= H\sgn{L} \times H\sgn{S}$ that is a direct product of the spaces of measured qudits. The qudits with Hilbert spaces $ H\sgn{L}=C^{N_L} $, $ H\sgn{S}=C^{N_S} $ have for each observable, $ A\sgn{L} $ and  $ A\sgn{S} $,  resolutions of identity \eqref{MatrEigen} denoted here as $\mathfrak{R}\left[A\sgn{L} \right] $ and  $\mathfrak{R}\left[A\sgn{S} \right] $: 
\begin{equation}\label{PrtRLS}
	\begin{array}{l}
		\mathfrak{R}\left[A\sgn{L} \right] = \left\lbrace \Pi_n \left[A\sgn{L} \right] =\roa{n;A\sgn{L}},\  n \in \left[0,N_L-1 \right]   \right\rbrace\\
		\mathfrak{R}\left[A\sgn{S} \right]= \left\lbrace  \Pi_p\left[A\sgn{S} \right]  =\roa{p;A\sgn{S}},\   p \in \left[0,N_S-1 \right]   \right\rbrace
	\end{array}
	.
\end{equation} 
The direct products of those produce $ \frac{N!}{N_L!\ N_S!} $  variants of composite qudit resolution of identity  $ \mathfrak{R}=\mathfrak{R}\left[A\sgn{L} \right]\times\mathfrak{R}\left[A\sgn{S} \right] $ that differ in the rules of composing the states of particles into composite states.  The rule to be applied here is:
\begin{equation}\label{PrtHash}
	r\left( n,p\right) =n+p  N_L,\quad n\in \left[0,N_L-1 \right],\ p \in \left[0,N_S-1 \right].
\end{equation} 
The other rules are formed through permutation of numbers of the states of composite qudit and the qudits -- particles.

Induced resolution of identity $  \mathfrak{R}$ and induced basis  $ \mathfrak{b} $ of a composite qudit, 
\begin{equation}\label{PrtRC}
	\begin{array}{l}
		\mathfrak{R}\left[A\sgn{L},A\sgn{S} \right]=\left\lbrace \Pi_{r\left( n,p\right)}\left[A\sgn{L},A\sgn{S} \right]  =\Pi_n\left[A\sgn{L} \right]\times \Pi_p\left[A\sgn{S} \right] ,\ \forall n,p   \right\rbrace\\
		\mathfrak{b}\left[A\sgn{L},A\sgn{S} \right]=\left\lbrace \cat{r\left( n,p\right);A\sgn{L},A\sgn{S}}  =\cat{n;A\sgn{L}}\times \cat{p;A\sgn{S}},\ \forall n,p  \right\rbrace
	\end{array},
\end{equation} 
for each rule $ r\left( n,p\right) $, determine the representation of the observables of particles by the observables of the composite. 

The basis of matrices of a composite qudit is formed by the direct products of the matrix bases of qudits -- particles
\begin{equation}\label{PrtBeta}
	\begin{array}{ll}
		\beta_{r\left(0 ,0\right)}=\hat{I}=\hat{I}\sgn{L}\times\hat{I}\sgn{S};\\
		\beta_{r\left(n ,0\right)}=B_n\sgn{L}=\beta_n\sgn{L}\times\hat{I}\sgn{S};&n \in \left[ 1, N^2_L-1\right]\\
		\beta_{r\left(0 ,p\right)}=B_p\sgn{S}=\hat{I}\sgn{L}\times\beta_p\sgn{S};& p \in \left[ 1, N^2_S-1\right]\\
		\beta_{r\left(n ,p\right)}=B_{n,p}=\beta_n\sgn{L}\times\beta_p\sgn{S} = B_n\sgn{L}B_p\sgn{S};
	\end{array}.
\end{equation}
It is obvious that all the $N^2=N^2_LN^2_S$ matrices are orthogonal due to orthogonality of matrices $\beta_n\sgn{L} $  and  $\beta_p\sgn{S} $ of the particle bases.

\subsubsection{Decomposition of a density matrix}

Here it is shown that a qudit state with the Hilbert space dimension $ N=N_LN_S $ can be represented by a composition of the particle states with Hilbert spaces $ N_L$ and $N_S $ in such a way that the direct products of resolutions of identity of the particles form the qudit state resolution of identity  $ \mathfrak{R}\left[\rho \right] =\mathfrak{R}\left[\rho\sgn{L} \right]\times\mathfrak{R}\left[\rho\sgn{S} \right] $ for the composite qudit, and the separating measurements for those particles produce the separating measurement for the composite qudit.

The composite qudit resolution of identity is the sequence \eqref{MatrEigen} of projectors on pure states
\[ \mathfrak{R}=\left\lbrace \Pi_r,\ r\in \left[0,N-1 \right]  \right\rbrace , \]
it can be reindexed with application of double indexation \eqref{PrtHash}:
\begin{equation}\label{PrtRN}
	\mathfrak{R}=\left\lbrace \Pi_{r(n,p)},\ n\in \left[0,N_L-1 \right], p\in \left[0,N_S-1 \right] \right\rbrace .
\end{equation}
Let the two sets  $\mathfrak{R}\sgn{L} $ and $ \mathfrak{R}\sgn{S}$ of compatible projectors are
\begin{equation}\label{PrtRtoLS}
	\begin{array}{l}
		\mathfrak{R}\sgn{L} = \left\lbrace \Pi_n\sgn{L}  =\sum_{p=0}^{N_S-1}\Pi_{r(n,p)},\  n \in \left[0,N_L-1 \right]   \right\rbrace\\
		\mathfrak{R}\sgn{S} = \left\lbrace  \Pi_p\sgn{S}  =\sum_{n=0}^{N_L-1}\Pi_{r(n,p)},\   p \in \left[0,N_S-1 \right]   \right\rbrace
	\end{array}
	.
\end{equation} 
Those projectors differ from the projectors of resolution of identity for separate qudits by the fact that the image of each projector $\Pi_n\sgn{L} $ from $ \mathfrak{R}\sgn{L} $ is a subspace with dimension $ N_S $, and for each projector $\Pi_p\sgn{S}$ from $\mathfrak{R}\sgn{S}  $ the image is a subspace with dimension $ N_L $.

The product of a pair of projectors $ \Pi_n\sgn{L}$  and  $\Pi_p\sgn{S} $,
\[ \Pi_n\sgn{L}\Pi_p\sgn{S}=\sum_{p'=0}^{N_S-1}\sum_{n'=0}^{N_L-1}\Pi_{r(n,p')}\Pi_{r(n',p)}= \Pi_{r(n,p)},\]
is equal to the respective projector $ \Pi_{r(n,p)} $  of resolution of identity \eqref{PrtRN}, since in the double sum only one term with indices $r(n,p') = r(n',p)\mapsto n=n'\ \&\ p=p'  $ is nonzero.

The sets of projectors \eqref{PrtRtoLS} are expanded to bases of local algebras. 

Let us denote
\begin{equation}\label{PrtBtoLS}
	\begin{array}{l}
		B\sgn{n,n'|L}_{x/y} =\sum_{p=0}^{N_S-1}B^{r(n,p),r(n',p)}_{x/y},\  n<n' \in \left[0,N_L-1 \right]   \\
		B\sgn{p,p'|S}_{x/y} =\sum_{n=0}^{N_L-1}B^{r(n,p),r(n,p')}_{x/y},\   p<p' \in \left[0,N_S-1 \right]   
	\end{array}
	.
\end{equation} 
The Hermitian $ N\times N $ matrices  $ B\sgn{n,n'|L}_{x/y} $, together with projectors $ \Pi_n\sgn{L} $, in amount of  $N_L^2$, form the set $ \left\lbrace B_n\sgn{L;N} ,n\in\left[ 1, N_L^2-1 \right] \right\rbrace $ of matrices that is the representation of the basis \eqref{Gbasis} of Hermitian $ N_L\times N_L $ matrices of qudit with dimension $ N_L $. Similarly, $ N\times N $  matrices $ B\sgn{p,p'|S}_{x/y}$, together with projectors  $\Pi_p\sgn{S} $, form the set $\left\lbrace B_p\sgn{S;N} ,p\in\left[1, N_S^2-1\right] \right\rbrace $ of matrices that is the representation  of the basis \eqref{Gbasis} of Hermitian  $ N_S\times N_S $  matrices of qudit with dimension $ N_S $. The products of these two sets of matrices complete these sets up to a vector basis of a composite qudit algebra of observables. The linear combinations 
\begin{equation}\label{PrtlocLS}
	A\sgn{L|loc}=A_0\hat{I}+\sum\limits_{n=1}^{N_L^2-1 }A_n\sgn{L}B_n\sgn{L;N},\quad
	A\sgn{S|loc}=A_0\hat{I}+\sum\limits_{p=1}^{N_S^2-1 }A_p\sgn{S}B_p\sgn{S;N}
\end{equation} 
form the subalgebras of local observables of the qudits $L $ and $S $ that are parts of the composite qudit algebra of observables. Combinations of basis matrices \eqref{Gbasis} with same coefficients $A_n\sgn{L} $ and   $A_p\sgn{S} $ 
\begin{equation}\label{PrtLSpart}
	A\sgn{L}=A_0\hat{I}\sgn{L}+\sum\limits_{n=1}^{N_L^2-1 }A_n\sgn{L}B_n\sgn{L},\quad
	A\sgn{S}=A_0\hat{I}\sgn{S}+\sum\limits_{p=1}^{N_S^2-1 }A_p\sgn{S}B_p\sgn{S}
\end{equation} 
are matrices $ N_L\times N_L $ and $ N_S\times N_S $ of the observables of the qudits -- particles. In fact difference between the local observables and the observables of the particles is not very significant, since those observables differ only with universal multipliers
\begin{equation}\label{PrtPartLoc}
	A\sgn{L|loc}=A\sgn{L}\times\hat{I}\sgn{S},\quad
	A\sgn{S|loc}=\hat{I}\sgn{L}\times A\sgn{S},
\end{equation} 
thus hereinafter the local observables \eqref{PrtlocLS} and the observables of the particles \eqref{PrtLSpart} are treated as local.

Joint measurement for the pair of observables \eqref{PrtRtoLS} with resolutions of identity $\mathfrak{R}\sgn{L} $ and $ \mathfrak{R}\sgn{S}$ is separating for the state of composite qudit with density matrix that has resolution of identity \eqref{PrtRN}.

Thus, for each state of composite qudit there exists such a pair of qudits -- particles that the separating measurement for the particles provides a separating measurement of the composite qudit state.

\subsubsection{  Density matrices}

The density matrix of a state in basis  \eqref{PrtBeta} is
\begin{equation}\label{PrtRho}
	\begin{array}{r}
		\hat{\rho}=\frac{1}{N}\hat{I} +\sum\limits_{n=1}^{N_L^2-1}\frac{d\sgn{L}_n}{M_nN_S}B_n\sgn{L}  +\sum\limits_{p=1}^{N_S^2-1}\frac{d\sgn{S}_p}{N_LM_p}B_p\sgn{S}\\
		+\sum\limits_{n=1}^{N_L^2-1}\sum\limits_{p=1}^{N_S^2-1}\frac{d_{n,p}}{M_nM_p}B_n\sgn{L}B_p\sgn{S}.
	\end{array}
\end{equation} 

The first term is fixed by the condition of normalization of the density matrix, the last term is responsible  for the mutual moments of the pairs of local observables, the second and the third terms  determine the density matrices of the particles
\begin{equation}\label{PrtRhoLoc}
	\rho\sgn{L}=\frac{1}{N_L}\hat{I}+\sum_{n=1}^{N_L^2-1}\frac{d\sgn{L}_n}{M\sgn{L}_n}B_n\sgn{L},\quad
	\rho\sgn{S}=\frac{1}{N_S}\hat{I}+\sum_{p=1}^{N_S^2-1}\frac{d\sgn{S}_p}{M\sgn{S}_p}B_p\sgn{S}
	.
\end{equation} 
The density matrices of the particles give for the observables of the particles average values equal to those given by the density matrix of the composite qudit  \eqref{PrtRho}.

Coefficients in \eqref{PrtRho} are equal to mathematical expectations for the observables represented by corresponding generalized Pauli matrices
\begin{equation}\label{PrtParam}
	d\sgn{L}_n=\aver{B_n\sgn{L}},\quad
	d\sgn{S}_p=\aver{B_p\sgn{S}},\quad
	d_{n,p}=\aver{B_n\sgn{L}B_p\sgn{S}}.
\end{equation}
Estimation of these coefficients can be obtained by the results of measurements for all combinations of generalized Pauli matrices forming the composite qudit or from the results of calculations. Determination of all the values \eqref{PrtParam} is a necessary and sufficient condition for complete determination of the composite qudit state.

\subsubsection{ Measurement of local observables}

Local measurement for one of the particles is performed by means of a device represented by a degenerate observable  $A\sgn{dg}$  of the composite qudit. It is accompanied by transformation of qudit state to subspace corresponding to eigenvalue $A\sgn{dg}_n$ registered in a specific measurement event number $ k $. Transformed density matrix is determined \cite{Lueders1951} by a normalized result of measured state density matrix $ \rho\sgn{m} $ wrapping by projectors $ \Pi_n\left[ A\sgn{dg}\right]  $ to that subspace,
\begin{equation}\label{PrtLocRed}
	\rho\sgn{m|r}_n=\frac{\Pi_n\left[ A\sgn{dg}\right]\rho\sgn{m} \Pi_n\left[ A\sgn{dg}\right] }{\mess{n;A\sgn{dg}}{\rho\sgn{m}}}.
\end{equation}

Measurement of the local observables of the composite qudit particles with resolutions of identity \eqref{PrtRLS} produces for separate events of measurement reduced density matrices  $\hat{\rho}\sgn{red|L}_n=\Pi_{n}\sgn{L}\times\hat{\rho}\sgn{S|L_n}$ and $\hat{\rho}\sgn{red|S}_p=\hat{\rho}\sgn{L|S_p}\times\Pi_{p}\sgn{S}$,  where conditional  reduced density matrices $\hat{\rho}\sgn{S|L_n}$ and $\hat{\rho}\sgn{L|S_p}$ are
\begin{equation}\label{PrtRLSred}
	\begin{array}{l}
		\hat{\rho}\sgn{S|L_n}= \frac{1}{N_S}\hat{I}
		+ \sum\limits_{p=1}^{N_S^2-1}\frac{\frac{d\sgn{S}_p}{N_LM_p}
			+\sum\limits_{s=1}^{N_L-1}\frac{d_{s,p}}{M_sM_p}C^{-1}_{s,n}}{\frac{1}{N_L}+\sum\limits_{s=1}^{N_L-1}\frac{d\sgn{L}_s}{M_s}C^{-1}_{s,n}}B_p\sgn{S};\\
		\hat{\rho}\sgn{L|S_p}= \frac{1}{N_L}\hat{I}
		+ \sum\limits_{n=1}^{N_L^2-1}\frac{\frac{d\sgn{L}_n}{N_SM_n}
			+\sum\limits_{r=1}^{N_S-1}\frac{d_{r,n}}{M_rM_n}C^{-1}_{r,p}}{\frac{1}{N_S}+\sum\limits_{r=1}^{N_S-1}\frac{d\sgn{S}_r}{M_r}C^{-1}_{r,p}}B_n\sgn{L}.\\
	\end{array}
\end{equation} 
The event of measurement for a qudit $ L $ local observable, with result $ n $, is accompanied by the composite qudit state reduction to density matrix $ \hat{\rho}\sgn{red|L}_n $ that is a direct product of the projector on the qudit $ L $  pure state $ \Pi_{n}\sgn{L} $ and the density matrix of the qudit $ S $ conditional state $ \hat{\rho}\sgn{S|L_n} $ . Likewise, the event of measurement for a qudit $ S $ local observable, with result $ p $, is accompanied by the composite qudit state reduction to density matrix $ \hat{\rho}\sgn{red|S}_p $ that is a direct product of the projector on the qudit $ S $ pure state $ \Pi_{p}\sgn{S} $ and the density matrix of the qudit $ L $ conditional state $ \hat{\rho}\sgn{L|S_p} $ .

Thus, the measurement of the local observable for one particle is accompanied by the composite qudit state reduction to the direct product of the pure state density matrix of the measured particle and the conditional state density matrix of the other particle. In general case the conditional state density matrices of one qudit that correspond to different pure states of the other one have different resolutions of identity; in more detail the properties of conditional states are characterised by the conditional phase portraits.

\subsubsection{Phase portrait of a composite qudit}
\label{DF}
The phase portrait of a composite qudit is determined by mathematical expectation
\begin{equation}\label{PrtPair}
	p\left(\psi,\phi \right) =\tr{\rho\roa{\psi}\times\roa{\phi}}=\mess{\psi}{\mess{\phi}{\rho}}
\end{equation}
of the product $ \Pi\left[ \psi,\phi\right] = \roa{\psi}\times \roa{\phi}$ of the particle $ L $ and $ S $ projectors to pure states, $ \Pi\sgn{L}\left[\psi \right]=\roa{\psi}\times\hat{I}\sgn{S}  $ and  $ \Pi\sgn{S}\left[\phi \right]=\hat{I}\sgn{L}\times  \roa{\phi}$ . 

The matrix elements of basis matrices
\begin{equation}\label{PrtVect}
	m\sgn{L}_n\left[\psi \right] =\mess{\psi}{\beta_n\sgn{L}},\quad
	m\sgn{S}_p\left[\phi \right]=\mess{\phi}{\beta_p\sgn{S}}
\end{equation}
form in the spaces $R^{N_L^2-1} $ and $R^{N_S^2-1} $ the vectors characterising the basis matrices of the qudits -- particles. Those vectors  produce the following expressions for the phase portrait of the composite 
\begin{equation}\label{PrtLS}
	\begin{array}{r}
		p\left(\psi,\phi \right) =		\frac{1}{N}+\sum\limits_{n=1}^{N_L^2-1}\frac{d\sgn{L}_nm\sgn{L}_n\left[\psi \right]}{M\sgn{L}_n} +\sum\limits_{p=1}^{N_S^2-1}\frac{d\sgn{S}_p}{M\sgn{S}_p}m\sgn{S}_p\left[\phi \right]\\
		+\sum\limits_{n=1}^{N_L^2-1}\sum\limits_{p=1}^{N_S^2-1}\frac{d_{n,p} m\sgn{L}_n\left[\psi \right]m\sgn{S}_p\left[\phi \right]}{M\sgn{L}_n M\sgn{S}_{p}},\\
	\end{array}
\end{equation}
the local phase portraits
\begin{equation}\label{PrtLSloc}
	\begin{array}{l}
		p\sgn{L}\left(\psi \right) =\frac{1}{N_L}
		+\sum\limits_{n=1}^{N_L^2-1}\frac{d\sgn{L}_nm\sgn{L}_n\left[\psi \right]}{M\sgn{L}_n },\\
		p\sgn{S}\left(\phi \right) =\frac{1}{N_S}
		+\sum\limits_{p=1}^{N_S^2-1}\frac{d\sgn{S}_pm\sgn{S}_p\left[\phi \right]}{M\sgn{S}_p }\\
	\end{array}
\end{equation}
and conditional phase portraits
\begin{equation}\label{PrtLScond}
	\begin{array}{l}
		p\sgn{L}\left(\psi |\phi\right) =\frac{1}{N_L}
		+\sum\limits_{p=1}^{N_L^2-1}\frac{d\sgn{L}_p\left[\phi \right]m\sgn{L}_n\left[\psi \right] }{M\sgn{L}_p },\\
		p\sgn{S}\left(\phi|\psi \right) =\frac{1}{N_S}
		+\sum\limits_{p=1}^{N_S^2-1}\frac{d\sgn{S}_p\left[\psi \right]m\sgn{S}_p\left[\phi \right]}{M\sgn{S}_p }\\
	\end{array}
\end{equation}
that differ by replacement of the parameters of the local density matrices $d\sgn{L}_p\mapsto d\sgn{L}_p\left[\phi \right]  $ $d\sgn{S}_p\mapsto d\sgn{S}_p\left[\psi \right]  $ with conditional parameters
\begin{equation}\label{PrtLSeff}
	d\sgn{L}_n\left[\phi \right]  =\frac{
		d\sgn{L}_n+
		\sum\limits_{p=1}^{N_S^2-1}\frac{d_{n,p}m\sgn{S}_p\left[\phi \right]}{ M\sgn{S}_{p}}
	}{p\sgn{S}\left(\phi \right)},\quad
	d\sgn{S}_p\left[\psi \right]  =\frac{
		d\sgn{S}_p+
		\sum\limits_{n=1}^{N_L^2-1}\frac{d_{n,p}m\sgn{L}_n\left[\psi \right]}{ M\sgn{L}_{n}}
	}{p\sgn{L}\left(\psi \right)}
	.
\end{equation}
The local and conditional phase portraits include scalar products of the vectors of state parameters with components $ d\sgn{S}_p $, $ d\sgn{L}_n $ or $ d\sgn{S}_p\left[\psi \right] $, $ d\sgn{L}_n\left[\phi \right] $ and vectors  $m\sgn{L}_n\left[\psi \right] $, $m\sgn{S}_p\left[\phi \right] $. In nonexceptional state of composite qudit the matrix $ \frac{d_{n,p}}{ M\sgn{L}_{p}} $ is nondegenerate, thus the direction of the conditional director of state for one particle changes with change of the direction of the counter of the other one, this is manifestation of entanglement.

Thus, the measured qudit state is split into states of the particles only by measuring devices matched with the state. Likewise, the measuring devices can split into particles only the states that are matched with the devices.

\subsection{Reduction with Separability or Entanglement}
\label{DR}

The measuring devices for which it is possible to make the measured state free of reduction provide the possibility of qudit separation into particles. For each state $\rho$ there exists a set of such devices $\mathcal{D}_{\rho}\subset \mathcal{D}$, and it is a measure zero subspace in the space of possible devices $\mathcal{D}$, thus reduction is a common result in measurement of composite qudit particles. 

Ability of a device with resolution of identity \eqref{PrtRC} to perform separation of a composite qudit in state $ \rho $ into particles with resolutions of identity \eqref{PrtRLS} is determined by the properties of the matrix \eqref{AtoRho} that transforms the resolution of identity for observables of the device to resolution of identity for the measured state. If this transformation is local, there exists a possibility to reconstruct the measuring devices for the particles in such a way that those devices begin to perform separating measurement of composite qudit. All the mentioned above states of composite qudit are splittable, measurement of the other composite qudit states by an arbitrary set of such devices providing possibility of qudit separation into particles is always followed by reduction.  

A set of states separated with a given measuring device is determined by local transformations forming a subgroup of the special unitary group $SU(N)$ of composite qudit transformations. This subgroup is a direct product of the special unitary groups of qudits -- particles $SU(N_L)$ and  $SU(N_S)$.  

The group   $SU(N)$  has as its elements matrices $ U\left[J \right]=\exp{iJ} $
with generators
\begin{equation}\label{PrtJ}
	J=\sum_{n=1}^{N_L^2-1}J\sgn{L}_nB_n\sgn{L}
	+\sum_{p=1}^{N_S^2-1}J\sgn{S}_pB_p\sgn{S}
	+\sum_{n=1}^{N_L^2-1}\sum_{p=1}^{N_S^2-1}J_{n,p}B_n\sgn{L}B_p\sgn{S}
\end{equation} 
that include sums of three types. The last one, the double sum, transforms the matrices of both particles jointly, while the first and the second sums generate the separate unitary transformations for the particles. Since the sets of matrices $B_n\sgn{L} $  and $ B_p\sgn{S}$ are complete sets of traceless matrices in respective spaces, the first and the second sums generate all the possible local transformations.

Not all of those transformations generate different measuring devices or different states. First of all this relates to the states differing by the sets of density matrix eigenvalues that have a common resolution of identity.  Those states have a common set of separable observables, thus it is rational to consider the transformation of resolution of identity  $ \mathfrak{R}\left[\rho \right]  $ only. Likewise, the measuring devices with common resolution of identity  $ \mathfrak{R}\left[A \right]  $ form a subalgebra of compatible observables.

There exist three classes of transformations of resolution of identity  $ \mathfrak{R}  $, those are the stabilizer subgroup $ U\sgn{st} $, the local subgroup $ U\sgn{loc}=SU(N_L)\times SU(N_S) $ and the entangling set $ U\sgn{ent}=SU(N_LN_S)\setminus (U\sgn{loc}\cup U\sgn{st} ) $ that is a quotient space of the group of composite qudit unitary transformations by the union of stabilizer and local subgroups. 
This classification leads to solution of the  local indistinguishability problem \cite{Ma2014,Walgate_2008,Walgate2000}.

\subsubsection{Stabilizer subgroup} 
Stabilizer subgroup $ U_{\mathfrak{R}} $ of resolution of identity  $ \mathfrak{R} $ consists of all the elements of the group for which this resolution of identity remains unchanged. Composite qudit resolution of identity \eqref{PrtRC} leaves unchanged all the elements of the group $SU(N)$, generators of which include in the expression \eqref{PrtJ} only diagonal elements, i.e. the terms of each sum with indices $n=n'\left(N_L+1 \right) $ and $p=p'\left(N_S+1 \right) $. From the definition of the basis matrices \eqref{Basis} it follows that the projectors of resolutions of identity for the particles are linear combinations of such elements, so the projectors commute with the elements of the stabilizer subgroup; the result of transformation for each projector is the initial projector $ U_{\mathfrak{R}}\Pi_k U_{\mathfrak{R}}^+=\Pi_k $. The diagonal elements are included to each of three sums in the expression \eqref{PrtJ} for the generator of an arbitrary element of the group. The stabilizer is the Abel subgroup with dimension $N-1=N_LN_S-1$. The conditions
\begin{equation}\label{ProStab}
	J\sgn{L}_n=0,\ J\sgn{S}_p=0,\  J_{n,p}=0,\quad\forall n\neq n'\left(N_L+1 \right) \ \& p \neq p'\left(N_S+1 \right)
\end{equation}
state that  $ U_{\mathfrak{R}} $ is a subsurface of the group $SU(N)$.

\subsubsection{Local subgroup} 
The subgroup of local transformations $ U_{\mathfrak{R}}\sgn{loc} $ of resolution of identity  $ \mathfrak{R} $ is determined by the conditions
\begin{equation}\label{ProStabC}
	J_{n,p}=0,\quad\forall \,n \& p ,
\end{equation}
coefficients $ J_{*} $ vanish for the products of local observables. 

The elements of two subgroups $SU(N_L)$ and $SU(N_S)$ are nontrivial transformations. Those subgroups commute, forming together the subgroup of local transformations of a composite qudit associated with the resolution of identity $ \mathfrak{R} $.  The generated by this subgroup set $ \mathfrak{B}\sgn{loc} $ of resolutions of identity is a surface with dimension $N_L^2-N_L+N_S^2-N_S $ in the space of resolutions of identity $ \mathfrak{B}_N $ with dimension $ N_L^2N_S^2-N_LN_S $.  This surface consists of resolutions of identity of all the observables measurement of which remains separating, its codimension $ \left(N_LN_S+N_L+N_S \right) \left(N_L-1 \right) \left(N_S-1 \right)  $ grows quadratically with growth of $ N_L $ and $ N_S $, thus the absence of entanglement of the results of measurements is a specific case.

By local transformations it is always possible to diagonalize the local density matrices \eqref{PrtRhoLoc}. From the physical point of view, existence of such transformations leads to a possibility to design local devices that provide separating measurement of local density matrices.

Let 
\[ U\sgn{L}=\exp{\left( i\sum_{n=1}^{N_L^2-1}J\sgn{L}_nB_n\sgn{L}\right) },\quad  U\sgn{S}=\exp{\left( i\sum_{p=1}^{N_S^2-1}J\sgn{S}_pB_p\sgn{S}\right) }\]
are the matrices of local transformations for the particles $L$ and $S$. Joint local transformation of the density matrix \eqref{PrtRho} leads to the matrix
\begin{equation}\label{PrtTr}
	\begin{array}{l}
		U\sgn{L}{U\sgn{S}}\hat{\rho}{U\sgn{L}}^+{U\sgn{S}}^+=\\
		\frac{1}{N}\hat{I}+U\sgn{L}\left( \sum\limits_{n=1}^{N_L^2-1}\frac{d\sgn{L}_n}{M_nN_S}B_n\sgn{L}\right) {U\sgn{L}}^+
		+{U\sgn{S}}\left( \sum\limits_{p=1}^{N_S^2-1}\frac{d\sgn{S}_p}{N_LM_p}B_p\sgn{S}\right) {U\sgn{S}}^+\\
		+\sum\limits_{n=1}^{N_L^2-1}\sum\limits_{p=1}^{N_S^2-1}\frac{d_{n,p}}{M_nM_p}\left( U\sgn{L}B_n\sgn{L}{U\sgn{L}}^+\right) \ \left( {U\sgn{S}}B_p\sgn{S}{U\sgn{S}}^+\right) .
	\end{array}
\end{equation}
It includes two matrices
\begin{equation}\label{PrtLocTr}
	D\sgn{L}=\sum_{n=1}^{N_L^2-1}\frac{d\sgn{L}_n}{M_nN_S}B_n,\quad  D\sgn{S}=\sum_{p=1}^{N_S^2-1}\frac{d\sgn{S}_p}{N_LM_p}B_p\sgn{S}
\end{equation}
that can be diagonalized by local transformations. If after this transformation the last sum in \eqref{PrtTr} includes only diagonal matrices $ B\sgn{L}_{n\left( N_L+1\right) } $ and $ B\sgn{S}_{p\left( N_S+1\right) } $, the transformed density matrix is separable.

Thus, the state of a composite qudit is separable if its density matrix commutes with the direct product of local density matrices.

\subsubsection{Entanglement}
The group  $SU(N)$ of transformations \eqref{AtoRho} of the resolution of identity for the observable to the resolution of identity for the measured state is a topological space with dimension $ N^2-N $ where the local transformations form a subspace with codimension $ \left(N_LN_S+N_L+N_S \right) \left(N_L-1 \right) \left(N_S-1 \right)  $   exceeding the dimension $N_L^2-N_L+N_S^2-N_S $  of this subspace of local transformations.

So, in general case the measuring device formed of the devices of the particles and the measured state are directed in such a way that the measurement entangles the contributions of the particles, and thus the properties of the entangling measurements have no features specific for entanglement.  
Consideration of the most simple examples of entangling measurements is given below.

The properties of the conditional phase portrait make it obvious that the entanglement is produced by the matrix of second moments (the last sum in \eqref{PrtRho}).

The entangling transformations have in the third sum  \eqref{PrtTr} terms with products of nondiagonal  elements of matrix bases of both particles. The most simple example with generator  $\sigma_{x}\sgn{01;L}\sigma_{x}\sgn{01;S}$ has a following shortcoming: without other generator terms, the local transformation (turn around the both axes, $y\sgn{L}$ and $y\sgn{S}$) turns this product to $\sigma_{z}\sgn{01;L}\sigma_{z}\sgn{01;S}$, and this one belongs to the stabilizer subgroup. Thus, a linear combination of at least two products is to be considered as a simplest entangling transformation; for instance, a linear combination $a\sigma_{x}\sgn{01;L}\sigma_{x}\sgn{01;S}+b\sigma_{y}\sgn{01;L}\sigma_{y}\sgn{01;S}$. This is split into a pair of combinations of ladder matrices \eqref{Ladder}
\begin{equation}\label{PrtEntSimple}
	\begin{array}{ll}
		J=  &\alpha a\sgn{01;L}{a\sgn{01;S}}^+ +\alpha^*{a\sgn{01;L}}^+{a\sgn{01;S}} \\
		+&\beta a\sgn{01;L}{a\sgn{01;S}} +\beta^*{a\sgn{01;L}}^+{a\sgn{01;S}}^+ .\\
	\end{array}
\end{equation}
The first pair corresponds to transfer of excitation from the mode 0 to the mode 1 for the particle $ S $, with reverse transfer of excitation for the particle $ L $ (the first term) and vice versa. The second one corresponds to increase or decrease of the excitation number of both particles, this is often treated as pair creation. Simple renumbering of the states of one particle can turn the first pair to another one and vice versa, so only one pair is worth attention.

The matrix of unitary transformation $ U\left[ J\right]=\exp{iJ}  $ generated by the first pair in \eqref{PrtEntSimple}, 
\begin{equation}\label{PrtEntSimpleU}
	\begin{array}{rl}
		U\left[ J\right]=&\hat{I}+\left( \cos\left|\alpha \right|-1\right)
		\left(\Pi\sgn{L}_0\Pi\sgn{S}_1 +\Pi\sgn{L}_1\Pi\sgn{S}_0 \right)\\
		+&i\sin \left|\alpha \right| \left( \frac{\alpha}{\left|\alpha \right|} a\sgn{01;L}{a\sgn{01;S}}^+ 
		+\frac{\alpha^*}{\left|\alpha \right|}{a\sgn{01;L}}^+{a\sgn{01;S}} \right),\\
	\end{array}
\end{equation}
transforms the separated states of composite qudit $ \cat{01}= \cat{0}\times\cat{1}  $ and $ \cat{10}= \cat{1}\times\cat{0}  $  to entangled ones
\begin{equation}\label{PrtEntSimpleCats}
	\begin{array}{rl}
		U\left[ J\right]\cat{01}=&\cos\left|\alpha \right|\cat{01} +i\sin \left|\alpha \right|\frac{\alpha^*}{\left|\alpha \right|} \cat{10}\\
		U\left[ J\right]\cat{10}=& i\sin \left|\alpha \right|\frac{\alpha}{\left|\alpha \right|}\cat{01} +\cos\left|\alpha \right|\cat{10}\\
	\end{array}
\end{equation}
and leaves unchanged all the other states of the separated basis of composite qudit.

Entangling transformations can change the states of all the pairs jointly, thus there can exist  $N_L N_S  $ variants of entanglement for a composite qudit. Classification of entangling transformations can be an object of a specific study.

\subsection{Qubit pair}
\label{DQ}
Entanglement is most evident in the measurement for a qubit pair. If the method for qubit separation is not of great value, such an object is as well called a ququart.

\subsubsection{Qubit pair spaces}

The space of a ququart pure states is a sphere $ S_6 $. The space of resolutions of identity for a ququart is $ \mathfrak{B}_4=S_6\times S_4\times S_2 $. It is transitive as to the effect of a  15 - parameter group $SU(4)$.

\begin{figure}[h!]	
	\centering
	\includegraphics[width=0.4\linewidth]{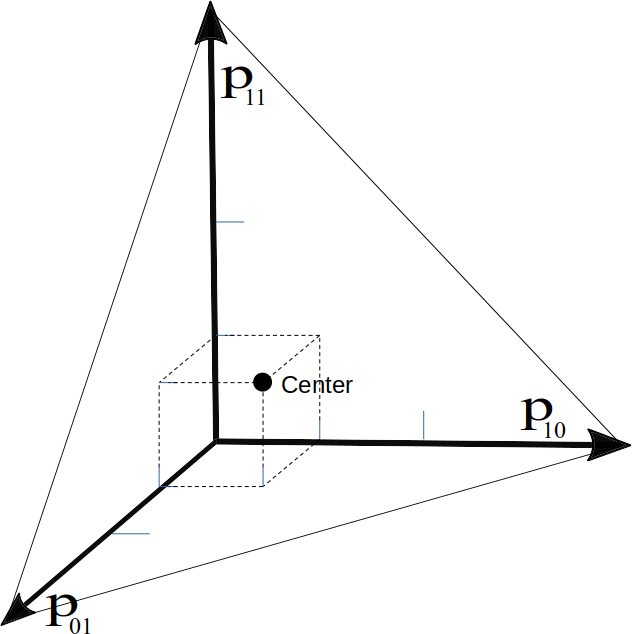}
	\vspace*{8pt}
	\caption{Simplex of density matrix eigenvalues for a pair of qubits is restricted by 4 planes given by conditions $ p_{00}\geq 0 $, $ p_{01}\geq 0 $, $ p_{10}\geq 0 $, $ p_{11}\geq 0 $ and the equation $ p_{00}+ p_{01}+p_{10}+p_{11}=1 $. In this graph  $ p_{01} $, $ p_{10} $, $ p_{11}$ are chosen as independent, thus $ p_{00} $ is determined by the equation $ p_{00}=1-p_{01} - p_{10} - p_{11} $.}\label{fig:symplex}
\end{figure}

The algebra of observables is a vector space $ R^{16} $  and is split into a product of the space of eigenvalues of observables $ R^{4} $ and the space of resolutions of identity  $ \mathfrak{B}_4 $. Likewise, the space of states is a direct product of the space of resolutions of identity $ \mathfrak{B}_4 $ and the simplex of density matrix eigenvalues $ T_3 $. This simplex is a trirectangular tetrahedron with vertices $ \left\lbrace\left[0,0,0 \right],\left[1,0,0 \right],\left[0,1,0 \right],\left[0,0,1 \right]  \right\rbrace  $, see figure \ref{fig:symplex}.

The coordinates of the simplex  $T_3$  point characterise the inputs of density matrix eigenstates to the mixed state represented by this point; this is similar to the case of a qubit mixed state where the input of two orthogonal pure states is characterised by the ratio of the sections of the diameter that goes through the point representing this state.
The tetrahedron vertices correspond to the pure states of the ququart, each face consists of the states for which the input of the pure state of the opposite vertex is absent, the edges correspond to the mix of two pure states. The centre of the simplex $T_3$ is in the point with coordinates $ p_{00}=p_{01} = p_{10} = p_{11} =1/4$.

\subsubsection{Density matrices}  
The basis of the ququart matrices is produced by 6 Pauli matrices
\begin{equation}\label{TwinQPaulies}
	\begin{array}{c}
		\Sigma_{x}\sgn{0}=\left( \begin{array}{rrrr}0&0&1&0\\0&0&0&1\\1&0&0&0\\0&1&0&0\\ \end{array}\right) ,\quad 
		\Sigma_{y}\sgn{0}=\left( \begin{array}{rrrr}0&0&-i&0\\0&0&0&-i\\i&0&0&0\\0&i&0&0\\ \end{array}\right) ,\quad 
		\Sigma_{z}\sgn{0}=\left( \begin{array}{rrrr}1&0&0&0\\0&1&0&0\\0&0&-1&0\\0&0&0&-1\\ \end{array}\right) , \\
		\Sigma_{x}\sgn{1}=\left( \begin{array}{rrrr}0&1&0&0\\1&0&0&0\\0&0&0&1\\0&0&1&0\\ \end{array}\right) ,\quad 
		\Sigma_{y}\sgn{1}=\left( \begin{array}{rrrr}0&-i&0&0\\i&0&0&0\\0&0&0&-i\\0&0&i&0\\ \end{array}\right) ,\quad 
		\Sigma_{z}\sgn{1}=\left( \begin{array}{rrrr}1&0&0&0\\0&-1&0&0\\0&0&1&0\\0&0&0&-1\\ \end{array}\right) . \\
	\end{array}
\end{equation} 
Three matrices $\Sigma_{a}\sgn{0}  $, like the three matrices  $\Sigma_{a}\sgn{1}  $, together with the unit matrix, produce  $ 4\times4 $ - representation of the qubit ordinary Pauli matrices. The products  $\Sigma_{ab}=\Sigma_{a}\sgn{0} \Sigma_{b}\sgn{1}  $ complete those matrices to the basis of the ququart matrix algebra.

The density matrix of an arbitrary state of a qubit pair is
\begin{equation}\label{PairRho}
	\hat{\rho}=\frac{1}{4}\hat{I}+\frac{1}{4}\vec{d}\sgn{0}\cdot\vec{\Sigma}\sgn{0}+\frac{1}{4}\vec{d}\sgn{1}\cdot\vec{\Sigma}\sgn{1}+\frac{1}{4}d_{ab}\Sigma_{ab}.
\end{equation}

The directors $\vec{d}\sgn{0} $ and $\vec{d}\sgn{1} $ are the directors of the local density matrices of qubits
\begin{equation}\label{PairRhoLoc}
	\hat{\rho}\sgn{0}=\frac{1}{2}\hat{I}+\frac{1}{2}\vec{d}\sgn{0}\cdot\vec{\sigma},\quad
	\hat{\rho}\sgn{1}=\frac{1}{2}\hat{I}+\frac{1}{2}\vec{d}\sgn{1}\cdot\vec{\sigma},
\end{equation}
as well.

Local transformations are performed by unitary matrices
\begin{equation}\label{PairULoc}
	\hat{U}\sgn{0}=\cos\varphi/2\hat{I}+\sin\varphi/2 \vec{\omega}\sgn{0}\cdot\vec{\Sigma}\sgn{0},\quad
	\hat{U}\sgn{1}=\cos\phi/2\hat{I}+\sin\phi/2 \vec{\omega}\sgn{1}\cdot\vec{\Sigma}\sgn{1}
\end{equation}
with the unit vectors $\vec{\omega}\sgn{0} $ and  $\vec{\omega}\sgn{1} $ determining the axes of rotation, and the angles of rotation $\varphi $ and  $\phi $ as parameters.

\subsubsection{Phase portrait of a qubit pair}

The phase portrait of a qubit pair is determined by the observable with the matrix
\begin{equation}\label{PairCount}
	\begin{array}{rl}
		\Pi\sgn{0}\left[\vec{m}\sgn{0}, \vec{m}\sgn{1}\right] =&\frac{1}{4}\left(\hat{I}+ \vec{m}\sgn{0}\cdot\vec{\Sigma}\sgn{0}\right) \left(\hat{I}+ \vec{m}\sgn{1}\cdot\vec{\Sigma}\sgn{1}\right)\\
		=&\frac{1}{4}\hat{I}+\frac{1}{4}\vec{m}\sgn{0}\cdot\vec{\Sigma}\sgn{0}+\frac{1}{4}\vec{m}\sgn{1}\cdot\vec{\Sigma}\sgn{1}+\frac{1}{4}m_{a}\sgn{0}m_{b}\sgn{1}\Sigma_{ab}.
	\end{array}
\end{equation}
The phase portrait of the state with density matrix \eqref{PairRho} is
\begin{equation}\label{PairP}
	P\left[\rho \right] \left( \vec{m}\sgn{0}, \vec{m}\sgn{1}\right) 
	=\frac{1}{4}+\frac{1}{4}\vec{m}\sgn{0}\cdot\vec{d}\sgn{0}+\frac{1}{4}\vec{m}\sgn{1}\cdot\vec{d}\sgn{1}+\frac{1}{4}m_{a}\sgn{0}m_{b}\sgn{1}d_{ab}
	.
\end{equation}
The local $P\sgn{0}\left[\rho \right] \left( \vec{m}\sgn{0}\right) $,  $ P\sgn{1}\left[\rho \right] \left( \vec{m}\sgn{1}\right)$ and conditional $P\sgn{0}_f\left[\rho \right] \left( \vec{m}\sgn{0}|\vec{m}\sgn{1}\right) $,  $P\sgn{1}_f\left[\rho \right] \left( \vec{m}\sgn{1}|\vec{m}\sgn{0}\right) $ phase portraits 
\begin{equation}\label{PairLocCond}
	\begin{array}{l}
		P\sgn{0}\left[\rho \right] \left( \vec{m}\sgn{0}\right) =
		\frac{1}{2}+\frac{1}{2}\vec{m}\sgn{0}\cdot\vec{d}\sgn{0},\\
		P\sgn{1}\left[\rho \right] \left( \vec{m}\sgn{1}\right) =
		\frac{1}{2}+\frac{1}{2}\vec{m}\sgn{1}\cdot\vec{d}\sgn{1},\\
		P\sgn{0}_f\left[\rho \right] \left( \vec{m}\sgn{0}|\vec{m}\sgn{1}\right) =
		\frac{1}{2}+\frac{1}{2}\vec{m}\sgn{0}\cdot\vec{d}\sgn{0f},\\
		P\sgn{1}_f\left[\rho \right] \left( \vec{m}\sgn{1}|\vec{m}\sgn{0}\right) =
		\frac{1}{2}+\frac{1}{2}\vec{m}\sgn{1}\cdot\vec{d}\sgn{1f}\\
	\end{array}
\end{equation}
are similar to the qubit phase portrait \eqref{PortraitQ}.

The effective directors, similarly to \eqref{PrtLSeff}, are 
\begin{equation}\label{PairDCond}
	d_a\sgn{0f}=\frac{d_a\sgn{0}+\sum d_{ab}m\sgn{1}_b}{1+\vec{m}\sgn{1}\cdot\vec{d}\sgn{1}},\quad
	d_a\sgn{1f}=\frac{d_a\sgn{1}+\sum d_{ab}m\sgn{0}_b}{1+\vec{m}\sgn{0}\cdot\vec{d}\sgn{0}}.
\end{equation}
Now it is obvious that entanglement of the results of measurement corresponds to a nondegenerate matrix of covariances $c_{ab}=d_{ab}-d_a\sgn{0}d_b\sgn{1}  $,  since for the effective director of the conditional phase portrait, $d_a\sgn{0f} $  or $ d_a\sgn{1f}$ , not only the value, but the direction as well depend on the direction of the counter, $ \vec{m}\sgn{1}$ or $\vec{m}\sgn{0} $. For the direction this dependence vanishes under the condition  $d_{ab}=k d_a\sgn{0} d_b\sgn{1} $, in this case there remains only the dependence of the value of the effective director of the conditional phase portrait on the direction of the counter 
\begin{equation}\label{PairND}
	d_{ab}=k d_a\sgn{0} d_b\sgn{1}\ \mapsto \quad	\begin{array}{l}
		
		\vec{d}\sgn{0f}=\frac{1+k\vec{m}\sgn{1}\cdot\vec{d}\sgn{1}}{1+\vec{m}\sgn{1}\cdot\vec{d}\sgn{1}}\vec{d}\sgn{0},\\
		\vec{d}\sgn{1f}=\frac{1+k\vec{m}\sgn{0}\cdot\vec{d}\sgn{0}}{1+\vec{m}\sgn{0}\cdot\vec{d}\sgn{0}}\vec{d}\sgn{1}.\\
	\end{array}
\end{equation}

Thus, absence of entanglement follows from a specific match between the measuring devices and the measured state.

Dependence of the direction of the effective director of a conditional phase portrait on the director of the counter is determined by the range of the  covariance matrix. Local transformations of the state vectors of both qubits bring the matrix $ c_{ab} $ to diagonal form
\begin{equation}\label{PairDdiag}
	c_{ab}= \sum_{k=1,2,3} C_k u_a\sgn{0|k}v_b\sgn{1|k},
\end{equation}
thus the range is equal  to the number of nonzero eigenvalues.

Three types of the dependence of the effective director $ \vec{d}\sgn{0f}\left(\vec{m}\sgn{1} \right)  $ of one qubit on the director of the counter of the other qubit $ \vec{m}\sgn{1} $ are possible.  In general case the range of the covariance matrix is 3, the effective director runs through all the possible directions, the entanglement is total. The covariance matrices of range 2 form a hypersurface, in this case the effective director varies only within the subspace formed by eigenvectors $u_a\sgn{0|1} $ and $u_a\sgn{0|2} $ of covariance matrix with nonzero eigenvalues  $ C_1 $ and   $ C_2 $, light entanglement is present.  If the range is 1 or 0, the effective director varies by the value only; such a behaviour is specific for classical correlation of observables, entanglement is absent.

\begin{figure}[h]
	\centering
	\includegraphics[width=0.3\linewidth]{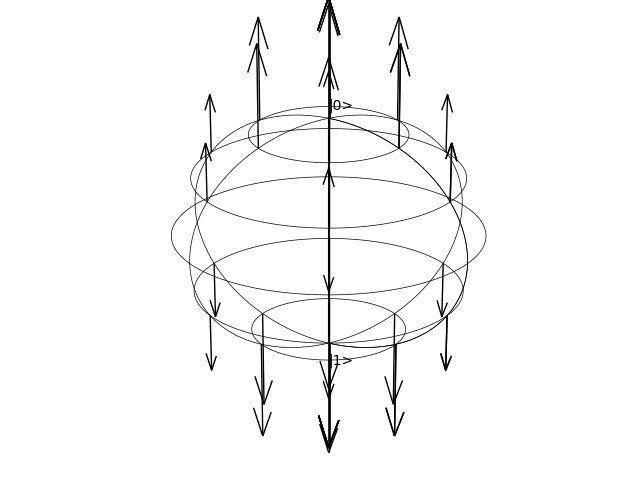}%
	\includegraphics[width=0.3\linewidth]{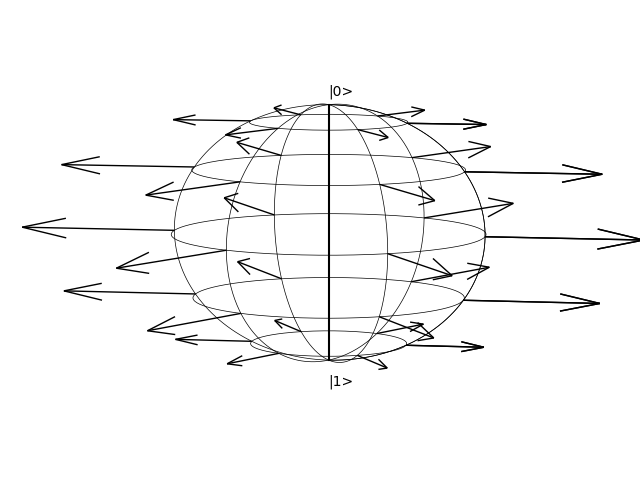}%
	\includegraphics[width=0.3\linewidth]{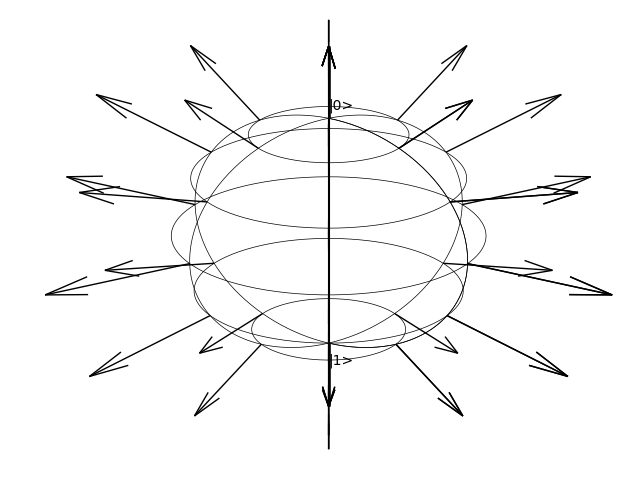}
	\vspace*{8pt}
	\caption{ Variants of the dependence of the qubit 0 effective director on the direction of the qubit 1 counter. 	 \\
		The effective director of the qubit 0 is represented by a vector with origin in the point of the sphere that is the terminal point of the director of the qubit 1 counter.\\
		Left -- the effective directors are coaxial; entanglement is absent.\\
		Centre -- the effective directors are normal to the axis of the sphere for all the counter directions; light entanglement.\\
		Right -- the effective directors coincide with the counter directions; total entanglement.\\
	}
	\label{fig:condprob}	
\end{figure}

Some examples of such a dependence are given in figure  \ref{fig:condprob} for a special case of zero directors of local density matrices; at that the covariance matrix is equal to the matrix of moments.

If all the three eigenvalues of this matrix are nonzero, the effective director of the qubit state 0 traces the direction of the counter of the qubit 1 (figure \ref{fig:condprob}, right), this corresponds to total entanglement of states; if only one eigenvalue of the matrix $ d_{ab} $ is zero, the effective director of state for the qubit 0 remains normal to the direction of the eigenvector $\vec{e}\sgn{0|3}$ of matrix $ d_{ab}$ for all the possible directions of the qubit 1 counter (figure \ref{fig:condprob}, centre), this corresponds to light entanglement of states; if at least two eigenvalues of the matrix $ d_{ab} $ are zero (figure  \ref{fig:condprob}, left), there is no entanglement.

\subsubsection{Light entanglement}
The entangled pure state of a qubit pair can always be reduced to canonical form that can be obtained from  \eqref{PrtEntSimpleCats} 
\[\cat{\psi}=\cos \psi\cat{01} +\sin \psi \cat{10},\quad \cat{\phi}=\cos \psi\cat{00} +\sin \psi \cat{11}.  \]
Entangling transformation of basis for a pair of qubits has similar form for each pair of vectors of entangled basis
\begin{equation}\label{PrtCanon}
	\begin{array}{ll}
		\cat{0,\left[\psi,\phi \right] }=\cos \psi\cat{00} +\sin \psi \cat{11},\\
		\cat{1,\left[\psi,\phi \right] }=\cos \phi\cat{01} +\sin \phi \cat{10},\\
		\cat{2,\left[\psi,\phi \right] }=-\sin \phi\cat{01} +\cos \phi \cat{10},\\
		\cat{3,\left[\psi,\phi \right] }=-\sin \psi\cat{00} +\cos \psi \cat{11}.\\
	\end{array}	
\end{equation}
It is split to two entangled pairs and has two independent parameters, $ \left[\psi,\phi \right] $. Thus, the measurement of the state of the system with density matrix
\[ \rho=\sum_{r=0}^3p_r\roa{r,\left[\psi,\phi \right]} \]
with counter projectors $ \Pi\sgn{0}=\roa{10}+\roa{11} $ and $ \Pi\sgn{1}=\roa{01}+\roa{11} $ can be separating if $ \left[\psi,\phi \right]=\left[0,0 \right] $ and lightly entangled if $ \left[\psi,\phi \right]=\left[\psi,0 \right] $ or  $ \left[\psi,\phi \right]=\left[0,\phi \right] $. Non-specific values of the parameters $ \left[\psi,\phi \right] $ correspond to total entanglement of the results of measurement for a nondegenerate mixed state.

\section{Multiqubit}
\label{DM}
Multiqubit  is a special type of qudit with a number of degrees of freedom that is equal to the power of two: $N=2^p$. The observables of the multiqubits are formed by the observables of $p$ independently measured qubits. 
The advantage of such formation of observables is in logarithmic decrease of the number of counters required for measurement of an observable of a multiqubit, since  $p$ pairs of counters produce in one act of measurement one of $N=2^p$ values with effectiveness similar to that for a composite $N$-dimension observable. It is even more effective in the case of $p$ counters of qubits completed with one common counter that registers the state orthogonal to the states registered by all those $p$ counters. This additional counter distinguishes the case of absence of response for all the qubit counters at registration of 0 for all the bits from a miss in measurement.

Measurement for a system of  $p$ qubits is performed by the counters of the qubits separately.
Mathematical representation of a qubit counter is given by one of the projectors of the pair \eqref{QAnorm} that forms the resolution of identity and is characterized by a point on the Bloch sphere or the director of the counter  $ \vec{m} $. The multiqubit counter is characterized by a set $ \mathcal{C}=\left\lbrace \vec{m}_q,\ q \in \left[0,p-1 \right]   \right\rbrace  $  of unit vectors -- the directors of the qubit counters. The set of possible values for the counter, i.e. the phase space of the multiqubit, is the direct product of Bloch spheres $ \mathfrak{P}\sgn{p}=\prod\limits_{q=0}^{p-1} S_2  $, the phase portrait of the multiqubit is the function of $p$ unit vectors: 
\begin{equation}\label{MPhPdef}
	P\left[\rho \right]\left( \vec{m}_0,\ldots\vec{m}_{p-1}\right) 
	=\tr{\rho\left( \Pi\sgn{0}\left(\vec{m}_{0} \right)\times\ldots\times \Pi\left(\vec{m}_{p-1} \right) \sgn{p-1}\right) }. 
\end{equation}

Since the vectors of the directors of completing projectors $\Pi\left(\vec{m} \right)  $,  $\overline{\Pi\left(\vec{\overline{m}} \right) } $ of the qubit resolution of identity have opposite directions $ \vec{\overline{m}}=-\vec{m} $, the indices of the system of projectors \eqref{MesPk} for the multiqubit resolution of identity have a $p$-bit representation: 
\begin{equation}\label{kb}
	k=b_0\cdot 2^{0}+b_{1}\cdot 2^{1}+\ldots+b_{p-1}\cdot 2^{p-1},\ b_q \in \left[ 0,1\right] .
\end{equation}
The point of phase space with coordinates $ M= \left\lbrace\vec{m}_{q},\ q\in \left[0,p-1 \right]   \right\rbrace  $ represents the multiqubit resolution of identity that consists of projectors
\begin{equation}\label{MulPset}
	\mathfrak{R}\left( M \right) =
	\left\lbrace \Pi_k\left( M \right),\ k\in \left[0,2^p-1 \right]   \right\rbrace, \quad
	\Pi_k\left( M \right)=
	\prod\limits_{q=0}^{p-1}\Pi\sgn{q}\left(\left(-1 \right)^{b_{q}\left(k \right) }\vec{m}_{q} \right).
\end{equation}
The phase portrait \eqref{MPhPdef} for this resolution of identity \eqref{MulPset} generates the probability distribution 
\[ \]
\begin{equation}\label{MultProbabDist}
	\begin{array}{l}
		\mathcal{P}\left( M \right)=\\
		\left\lbrace p_k= p\left[\rho \right]\left(\left(-1 \right)^{b_{0}\left(k \right) }\vec{m}_{0},\ldots ,\left(-1 \right)^{b_{p-1}\left(k \right) }\vec{m}_{p-1} \right),\ k\in \left[0,2^p-1 \right]   \right\rbrace.\\
	\end{array}
\end{equation}
The density matrix of a qubit in the multiqubit can be determined by supplementing the measurement by the qubit counter from resolution of identity \eqref{MulPset} with two incompatible counters. Hereinafter it is shown that the measurement of joint probability distributions \eqref{MultProbabDist} for all the combinations of three incompatible counters of each qubit is enough for determination of the multiqubit density matrix.

\subsection{Multiqubit decomposition}

Similarly to decomposition of a qudit to a pair of particles, a state of a multiqubit has a representation by such a composition of  $p$ qubits that the separating measurements of local states for each qubit produce a separating measurement for the multiqubit. Construction of such a representation starts from denoting
\begin{equation}\label{ktob}
	\mathbf{k}_q=\left\lbrace b_0+\ldots+0\cdot 2^{q}+\ldots+b_{p-1}\cdot 2^{p-1},\ b_m=0,1 \right\rbrace
\end{equation}
for the set of all the numbers within $ \left[0,2^{p-1} \right]  $ for which the $ q $-th bit of binary representation is zero. Each such set, with its complement, i.e. the set $ \overline{\mathbf{k}}_q=\left\lbrace k+2^{q},\ k \in  \mathbf{k}_q\right\rbrace  $ in which all the indices for which the $ q $-th bit is 1, cowers all the set of multiqubit indices. 

For each $ 0\leq q\leq p-1 $  three matrices
\begin{equation}\label{Qgenerators}
	\Pi\sgn{q}=\sum_{ k \in \mathbf{k}_q} \Pi_k,\quad 
	{A\sgn{q}}=\sum_{ k \in \mathbf{k}_q} \ro{k}{k+2^q},\quad 
	{A\sgn{q}}^+=\sum_{ k \in \mathbf{k}_q} \ro{k+2^q}{k}
\end{equation}
generate the subalgebra equivalent to the qubit algebra of observables.

Really, the elements of this subalgebra,
\begin{equation}\label{QPauli}
	\Sigma_1\sgn{q} =A\sgn{q}+{A\sgn{q}}^+,\quad
	\Sigma_2\sgn{q} =iA\sgn{q}-i{A\sgn{q}}^+,\quad
	\Sigma_3\sgn{q} =\hat{I}-2\Pi\sgn{q},
\end{equation}
with the products
\[ \Sigma_1\sgn{q}\Sigma_2\sgn{q}=i\Sigma_3\sgn{q},\quad  \Sigma_3\sgn{q}\Sigma_1\sgn{q}=i\Sigma_2\sgn{q},\quad  \Sigma_2\sgn{q}\Sigma_3\sgn{q}=i\Sigma_1\sgn{q},  \]
realize the representation of the Pauli matrix algebra with matrices $ N\times N $.

The ladder matrices ${A\sgn{q}}^+ $, ${A\sgn{q}}  $, with $\hat{I}-\Pi\sgn{q} ={A\sgn{q}}^+  {A\sgn{q}}$ , are to be interpreted as the creation/annihilation operators and the number operator for the qubit $ {q} $.

The products of the qubit projectors 
\begin{equation}\label{QgenRevers}
	\Pi_k=\prod_{q=0}^{p-1}\Pi\sgn{q}_{b_q\left( k\right) }=\prod_{q=0}^{p-1}\frac{\hat{I}+(-1)^{b_q\left( k\right)}\Sigma_3\sgn{q}}{2}
\end{equation}
for each number $k \in 0\ldots N-1$ with bit representation \eqref{kb} are the projectors of resolution of identity \eqref{MesPk} for the multiqubit. 

Measurement of the multiqubit projector $ \Pi_k $ is equivalent to joint measurement of the set of qubit projectors $\left\lbrace \Pi\sgn{q}_{b_q\left( k\right) },\ q \in \left[0,p-1 \right]   \right\rbrace $. For each state there exists a multiqubit decomposition into a system of qubits for which measurement does not lead to entanglement.

\subsection{Density Matrices} 
Hermitian matrices for multiqubit are represented by combinations of Pauli matrices. It is convenient to group the terms that include same numbers of Pauli matrices of different particles:
\begin{equation}\label{MulA}
	\begin{array}{rl}
		A=&A_0\hat{I}+\sum\limits_{q=0}^{p-1}\sum\limits_{a=1,2,3}A\sgn{q}_a\Sigma\sgn{q}_a
		+\sum\limits_{q'>q=0}^{p-1}\sum\limits_{a=1,2,3}A\sgn{qq'}_{a_q,a_{q'}}\Sigma\sgn{q}_{a_q}\Sigma\sgn{q'}_{a_{q'}}\\
		+&\ldots
		+\sum\limits_{a_0,\ldots,a_{p-1}=1,2,3}A_{a_0,\ldots,a_{p-1}}\Sigma\sgn{0}_{a_0}\ldots\Sigma\sgn{p-1}_{a_{p-1}}
	\end{array}.
\end{equation}
The expression for the multiqubit density matrix takes into account that the trace is equal to one and the square of each of the Pauli matrices \eqref{QPauli} is proportional to unity matrix:
\begin{equation}\label{MulRho}
	\begin{array}{rl}
		\rho=&\frac{1}{N}\hat{I}+\frac{1}{N}\sum\limits_{q=0}^{p-1}\sum\limits_{a=1,2,3}d\sgn{q}_a\Sigma\sgn{q}_a
		+\frac{1}{N}\sum\limits_{q'>q=0}^{p-1}\sum\limits_{a=1,2,3}d\sgn{qq'}_{a_q,a_{q'}}\Sigma\sgn{q}_{a_q}\Sigma\sgn{q'}_{a_{q'}}\\
		+&\ldots
		+\frac{1}{N}\sum\limits_{a_0,\ldots,a_{p-1}=1,2,3}d_{a_0,\ldots,a_{p-1}}\Sigma\sgn{0}_{a_0}\ldots\Sigma\sgn{p-1}_{a_{p-1}}
	\end{array}.
\end{equation}
The parameters $ d\sgn{*}_{*} $ characterize deflection of density matrix from the equilibrium one and are equal to mathematical expectations for respective Pauli matrices or combinations of those,
\begin{equation}\label{MulRhoD}\begin{array}{lr}
		d\sgn{q}_a=\aver{\Sigma\sgn{q}_a},&d\sgn{qq'}_{a_q,a_{q'}}=\aver{\Sigma\sgn{q}_a\Sigma\sgn{q'}_{a_{q'}}},\\
		\ldots&d_{a_0,\ldots,a_{p-1}}=\aver{\Sigma\sgn{0}_{a_0}\ldots\Sigma\sgn{p-1}_{a_{p-1}}}.\\
	\end{array}
\end{equation}
Thus, to determine the multiqubit density matrix it is necessary and sufficient to determine the mathematical expectations for all the combinations of three Pauli matrices of each qubit.

The local density matrices of each qubit follow from \eqref{MulRho} :
\begin{equation}\label{MulRhoLoc}
	\rho\sgn{q}=\frac{1}{2}\hat{I}+\frac{1}{2}\sum\limits_{a=1,2,3}d\sgn{q}_a\Sigma\sgn{q}_a,
\end{equation}
the partial density matrix of a pair of qubits with numbers $ q$ and  $r $ is
\begin{equation}\label{MulRhoLocR}
	\begin{array}{l}
		\rho\sgn{q,r}=\\
		\frac{1}{4}\hat{I}+\frac{1}{4}\sum\limits_{a=1,2,3}d\sgn{q}_a\Sigma\sgn{q}_a
		+\frac{1}{4}\sum\limits_{a=1,2,3}d\sgn{r}_a\Sigma\sgn{r}_a
		+\frac{1}{4}\sum\limits_{a,b=1,2,3}d\sgn{qr}_{a,b}\Sigma\sgn{q}_{a}\Sigma\sgn{r}_{b}.
	\end{array}
\end{equation}
Here $ d\sgn{q}_a$, $ d\sgn{r}_a $ and  $ d\sgn{qr}_{a,b}$ are the coefficients of the density matrix \eqref{MulRho} with respective numbers.

Similarly, the partial density matrix of a set of $ s $ qubits is derived from  \eqref{MulRho} by a simple deletion of all the terms that include the Pauli matrices of qubits not included to the partial subsystem. Additionally,  the normalizing denominator $ N=2^p $ is replaced by $N_{part}= 2^s $.

\subsection{Phase Portrait}

The counter of an arbitrary pure state of a multiqubit is parametrized by a point $\mathfrak{m}= \left\lbrace \vec{m}\sgn{q},\ q\in \left[0,p-1 \right]\right\rbrace  $ on the phase space of multiqubit and is represented by the product of projectors on the qubit pure states \eqref{QAnorm}:  
\begin{equation}\label{MulPhProj}
	\Pi\left( \mathfrak{m} \right)=\prod\limits_{q=0}^{p-1}\Pi\sgn{q}\left( \vec{m}\sgn{q} \right) =\frac{1}{2^p}\prod\limits_{q=0}^{p-1}\left(\hat{I}+\vec{m}\sgn{q}\vec{\Sigma}\sgn{q} \right). 
\end{equation}
Mathematical expectation for this counter with density matrix \eqref{MulRho} is a phase portrait of the multiqubit state, 
\begin{equation}\label{MultProbab}
	p\left[\rho \right]\left(\mathfrak{m}\right)  = \frac{1}{N}+
	\frac{1}{N}\sum_{q=0}^{p-1} d\sgn{q}_{a}m\sgn{q}_{a} 
	+\ldots \frac{d_{a_0\ldots a_{p-1}}}{N}m\sgn{0}_{a_0}\ldots m\sgn{p-1}_{a_{p-1}}.
\end{equation}
The phase portrait is a linear function of the counter director for each qubit $ \vec{m}\sgn{q} $. For each counter the completing projector $\overline{\Pi}\sgn{q}\left( \vec{m}\sgn{q} \right)=\Pi\sgn{q}\left(- \vec{m}\sgn{q} \right)$ differs by the sign of the director only, thus adding of the inputs from the projector with director $ \vec{m}\sgn{q} $ and from its complement with director $ -\vec{m}\sgn{q} $ is equivalent to deletion of all the terms that include the counter director of respective qubit, with doubling the result.

The local phase portrait of a qubit $ q $ is generated by its density matrix
\begin{equation}\label{MultPLoc}
	p\sgn{q}\left(\vec{m} \right)=\frac{1}{2}+\frac{1}{2}\vec{d}\sgn{q}\cdot\vec{m}.
\end{equation}
It depends only on the director of the local counter  $ \vec{m} $ and is determined by the value and direction $ \vec{d}\sgn{q} $ of deflection of the local state from the equilibrium one. 

The conditional local phase portrait of a selected qubit, let it is the qubit 0, is determined by all the set of coordinates of the point on the phase space of the multiqubit:
\begin{equation}\label{MultPCond}
	p_0\sgn{cond}\left(\vec{m}|\mathfrak{m}\sgn{0}\right)=\frac{1}{2}+\frac{1}{2}\vec{d}\sgn{eff}\left(\mathfrak{m}\sgn{0} \right)\cdot\vec{m},\quad \mathfrak{m}\sgn{0}=\vec{m}\sgn{1}\ldots\vec{m}\sgn{p-1}.
\end{equation}
The effective director $\vec{d}\sgn{eff}\left(\mathfrak{m}\sgn{0} \right)  $ is given by the expression
\begin{equation}\label{MultDCond}
	d_a\sgn{eff}\left(\mathfrak{m}\sgn{0} \right)= \frac
	{d\sgn{0}_a+\sum\limits_{q=1}^{p-1}d\sgn{0q}_{ab}m\sgn{q}_{b}
		+\ldots+d_{a,a_1,\ldots,a_{p-1}}m\sgn{1}_{a_1}\ldots m\sgn{p-1}_{a_{p-1}}}
	{1+\sum\limits_{q=1}^{p-1} d\sgn{q}_{a}m\sgn{q}_{a}	+\ldots  + d\sgn{1\ldots,p-1}_{a_1,\ldots,a_{p-1}}m\sgn{1}_{a_1}\ldots m\sgn{p-1}_{a_{p-1}}}.
\end{equation}
The entanglement of the results of measurement for the state of multiqubit has the form of the dependence of direction of the conditional phase portrait effective director on the directions of the counters of the other qubits. For the state with nondegenerate density matrix all the parameters $ d\sgn{*}_{*} $ are nonzero, thus the entanglement is a general property of multiqubit measurement.

The specific cases of partial entanglement are possible if a sufficient number of parameters $ d\sgn{*}_{*} $ turns to zero.

\subsection{ Reduction and Entanglement}
Similarly to the case of measurement for a pair of qudits, the measurement of a multiqubit can be separating, or can be accompanied by reduction with or without entanglement of the results of the measurement for qubits. The type of the measurement is determined by the properties of the matrix  $ U\left[J \right]=e^{i J} $  of transformation \eqref{AtoRho} of resolution of identity for the observable to resolution of identity for the measured state density matrix. The generator  $J $ of transformation matrix is a Hermitian matrix with trace  0 and is expressed by 
\begin{equation}\label{MulJ}
	J=\sum\limits_{q=0}^{p-1}J\sgn{q}_a\Sigma\sgn{q}_a
	+\sum\limits_{q'>q=0}^{p-1}J\sgn{qq'}_{a_q,a_{q'}}\Sigma\sgn{q}_{a_q}\Sigma\sgn{q'}_{a_{q'}}
	+
	\ldots
	+J_{a_0,\ldots,a_{p-1}}\Sigma\sgn{0}_{a_0}\ldots\Sigma\sgn{p-1}_{a_{p-1}}.
\end{equation}

Depending on what terms are present or absent in the above expression, transformation can be identical, local or entangling. 

\textbf{Transformation -- stabilizer} has a matrix that does not change the projectors \eqref{MulPset},
\[ U\left[ J_s\right]\Pi_kU\left[ J_s\right]^+= \Pi_k. \]
It is produced by generator
\begin{equation}\label{MulStab}
	J_s=\sum\limits_{q=0}^{p-1}J\sgn{q}_3\Sigma\sgn{q}_3+\ldots+J_m\Sigma\sgn{0}_{3}\ldots\Sigma\sgn{p-1}_{3}
\end{equation}
that includes only the matrices $ \Sigma\sgn{q}_{3} $ for each qubit. Resolutions of identity for the measuring device and the measured state are the same, thus the measurement is separating. In topological space $SU(2^p)$ with real dimension $ 2^{2p}-1=N^2-1 $ the subgroup -- stabilizer forms a subspace with dimension $ 2^{p}-1=N-1 $, so it is a set with measure null.

\textbf{Local transformation} is produced by a linear by Pauli matrices generator,
\begin{equation}\label{MulJloc}
	\begin{array}{l}
		J\sgn{loc}=\sum\limits_{q=0}^{p-1}j\vec{n}_{q}\vec{\Sigma}\sgn{q},\mapsto U\left[ J\right] =\prod\limits_{q=0}^{p-1}U\left[ j\vec{n}_{q}\vec{\Sigma}\sgn{q}\right],\\
		U\left[ j\vec{n}\sgn{q}\vec{\Sigma}\sgn{q}\right]=\cos j\sgn{q}\ \hat{I}+i\sin j\sgn{q}\ \vec{n}_q\vec{\Sigma}\sgn{q},
	\end{array}	
\end{equation}
and is split to the product of transformations of qubits with matrices \eqref{UexpJ}; reduction of states takes place for each qubit separately, so the measurement is local.

Only local transformations generated by matrices  $  \Sigma_{1|2}\sgn{q}$ of each qubit differ from the elements of the stabilizer, those transformations form the subspace with dimension $2^p=N$.

\textbf{Entangling transformations.}   All the transformations of the resolution of identity for the observable to the resolution of identity for the density matrix of the measured state that are not included to the subgroup -- stabilizer or to the subgroup of local transformations entangle the results of measurement for the qubits of the multiqubit. Those transformations form a subspace with dimension $N^2-1-\left(N-1 \right) -N=N^2-2N=2^{2p}-2^{p+1}$.

Classification of the variants of entanglement can be an object of a specific study; hereinafter only maximal entanglement is considered, i.e. the case of the terms with Pauli matrices $ \Sigma\sgn{q}_{x} $, $ \Sigma\sgn{q}_{y} $ of each qubit being present in the generator.

Maximal entanglement is arranged similarly for the multiqubits irrespectively of the number of qubits. Effect of Pauli matrices $\Sigma\sgn{q}_{x} $ and $ \Sigma\sgn{q}_{y} $ on not entangled basis vectors differs only by the phase multiplier,  
\[ \Sigma\sgn{q}_{x}\cat{\ldots,b_q,\ldots} =\cat{\ldots,\overline{b}_q,\ldots},\quad
\Sigma\sgn{q}_{y}\cat{\ldots,b_q,\ldots} =e^{\left(-1 \right)^{b_q} i\pi/2}\cat{\ldots,\overline{b}_q,\ldots},\]  
the matrix $ \Sigma\sgn{q}_{x} $ inverts the value of respective bit only, $ b_q\mapsto \overline{b}_q $, while the matrix $ \Sigma\sgn{q}_{y} $ additionally changes the phase. Inversion of all the bits in the index of the basis vector is equivalent to replacement $ k\mapsto 2^p-k-1 $, thus an arbitrary generator of maximally entangling transformation  $ J\sgn{max} $  entangles the pairs of states $ \cat{k} $ and  $ \cat{2^p-k-1} $,
\[J\sgn{max}\cat{k} =j_ke^{i\varphi_k} \cat{2^p-k-1},\quad 
J\sgn{max}\cat{2^p-k-1} =j_ke^{-i\varphi_k} \cat{k}, \]
and produces the transformation matrix $ U\left[J\sgn{max} \right] $,
\begin{equation}\label{MulJmax}
	U\left[J\sgn{max} \right]=\sum\limits_{k=0}^{2^p-1} \left(\cos j_k+i\frac{\sin{j_k}}{j_k}J\sgn{max} \right)\Pi_k .
\end{equation}

As to more complicated variants of entanglement of multiqubit, it is better to analyse those in representation by means of a conditional phase portrait.

\section{Conclusions}
Measuring devices are based on the counters for which the result of measurement is the number of responses, those counters perform physical implementation of von Neumann projective observables. Each nondegenerate observable is characterized by a respective set of eigenvalues and a resolution of identity. Resolution of identity is realized by a complete compatible system of counters of pure states.  Resolution of identity of a composite qudit is realized by a complete set of all the counters of pure states of the particles, with all possible combinations of counters.

The result of measurement of a qudit nondegenerate observable is determined by three groups of quantities, those are: the set of eigenvalues of the observable, the set of eigenvalues of the density matrix of the measured state and the matrix of transformation of resolution of identity for the observable to resolution of identity for the density matrix. If this transformation is equal to the identical one, the measurement is a separating measurement; otherwise it is accompanied by reduction of the measured state. The norm of generator of the unitary matrix of transformation of resolution of identity for the observable to resolution of identity for the density matrix can be used as the measure of reduction.

Decoding of information coded by a multiqubit or a qudit state is possible under condition of realization of a statistically valuable measurement series for $ N+1 $ incompatible observables. Inaccuracy of decoding is determined by variance of each observable.

Information transfer by a quantum channel is performed by a separable observable with accuracy up to the match between the resolutions of identity of the source and the receiver.

Physically meaningful are those degenerate observables that split the phase space to a product of the spaces of the particles in such a way that the products of the observables of the particles form a nondegenerate observable. The problem of reconstruction of a composite state by measurement of the observables of the particles is solved by determination of all the moments of the observables for the particles. 

Each set of observables of the particles can perform a separating measurement for a limited subset of composite states, the complement of this subset consists of entangled states. Each composite state has its own complex of observables of the particles that realises the separating measurement of the state. The set of those complexes is obtained by factorization of the set of resolutions of identity by the group of local transformations.

The qubit state is determined by mathematical expectations of three Pauli matrices that form a three-dimensional vector in space invariant with respect to the group of rotations $ O_3 $ induced by a unitary group $ SU(2) $ of the qubit Hilbert space. The direction of this vector determines the director of the non-demolition measurement counter, its length -- difference of probabilities for two variants of response of the counter. 

The multiqubit state is determined by mathematical expectations of the Pauli matrices of each qubit separately and in all possible combinations. The counter of nondegenerate measurement is characterized by the set of counter directors for each qubit. The set of mathematical expectations of the counter forms the multiqubit phase portrait, the local and the conditional phase portraits of the qubits.  The conditional phase portrait of each qubit is an efficient instrument for differentiation of separable states with local reduction and the entangled states with measurement reduction not removable by choosing local counters.


\begin{thebibliography}{10}
\bibitem{Schumacher95}
Benjamin Schumacher.
\newblock Quantum coding.
\newblock {\em Phys. Rev. A}, 51:2738--2747, Apr 1995.
\bibitem{shannon1}
C.~E. Shannon.
\newblock A mathematical theory of communication.
\newblock {\em Bell System Tech. J.}, 27:379--423\& 623--656, 1948.

\bibitem{vonneumann}
J~von Neumann.
\newblock {\em {Mathematische Grundlagen der Quantenmechanik}}.
\newblock Springer Verlag, Berlin, 1932.

\bibitem{nielsen}
M~A Nielsen and I~L Chuang.
\newblock {\em {Quantum Computation and Quantum Information}}.
\newblock Cambridge University Press, Cambridge, 2000.

\bibitem{WheelerZurek}
J~A Wheeler and W~H Zurek, editors.
\newblock {\em {Quantum Theory and Measurement}}.
\newblock Princeton University Press, 1983.

\bibitem{CERF1998}
Nicolas~J. Cerf and Chris Adami.
\newblock Information theory of quantum entanglement and measurement.
\newblock {\em Physica D: Nonlinear Phenomena}, 120(1):62--81, 1998.
\newblock Proceedings of the Fourth Workshop on Physics and Consumption.

\bibitem{ekert91}
Artur~K. Ekert.
\newblock Quantum cryptography based on bell's theorem.
\newblock {\em Phys. Rev. Lett.}, 67:661--663, Aug 1991.

\bibitem{Bennett96A}
C.~H. Bennett, H.~J. Bernstein, S.~Popescu, and B.~Schumacher.
\newblock Concentrating partial entanglement by local operations.
\newblock {\em Phys. Rev. A}, 53:2046, 1996.

\bibitem{PHPLA}
P.~Horodecki.
\newblock Separability criterion and inseparable mixed states with positive
partial transposition.
\newblock {\em Phys.Lett.A}, 232:333, 1997.

\bibitem{horodecki:865}
Ryszard Horodecki, Pawe\l Horodecki, Micha\l Horodecki, and Karol Horodecki.
\newblock Quantum entanglement.
\newblock {\em Reviews of Modern Physics}, 81(2):865, 2009.

\bibitem{Mancini202010}
Masoud Gharahi, Stefano Mancini, and Giorgio Ottaviani.
\newblock Fine-structure classification of multiqubit entanglement by algebraic
geometry.
\newblock {\em Phys. Rev. Research}, 2:043003, Oct 2020.

\bibitem{Martin201812}
A.~Neven, J.~Martin, and T.~Bastin.
\newblock Entanglement robustness against particle loss in multiqubit systems.
\newblock {\em Phys. Rev. A}, 98:062335, Dec 2018.

\bibitem{Guhne201807}
Nikolai Wyderka, Felix Huber, and Otfried G\"uhne.
\newblock Constraints on correlations in multiqubit systems.
\newblock {\em Phys. Rev. A}, 97:060101, Jun 2018.

\bibitem{Vedral201709}
Huangjun Zhu, Zhihao Ma, Zhu Cao, Shao-Ming Fei, and Vlatko Vedral.
\newblock Operational one-to-one mapping between coherence and entanglement
measures.
\newblock {\em Phys. Rev. A}, 96:032316, Sep 2017.

\bibitem{Adesso201608}
C\'ecilia Lancien, Sara Di~Martino, Marcus Huber, Marco Piani, Gerardo Adesso,
and Andreas Winter.
\newblock Should entanglement measures be monogamous or faithful?
\newblock {\em Phys. Rev. Lett.}, 117:060501, Aug 2016.

\bibitem{Misra201608}
Tamoghna Das, Sudipto~Singha Roy, Shrobona Bagchi, Avijit Misra, Aditi Sen(De),
and Ujjwal Sen.
\newblock Generalized geometric measure of entanglement for multiparty mixed
states.
\newblock {\em Phys. Rev. A}, 94:022336, Aug 2016.

\bibitem{Zukowski201112}
Wies\l{}aw Laskowski, Marcin Markiewicz, Tomasz Paterek, and Marek
\ifmmode~\dot{Z}\else \.{Z}\fi{}ukowski.
\newblock Correlation-tensor criteria for genuine multiqubit entanglement.
\newblock {\em Phys. Rev. A}, 84:062305, Dec 2011.

\bibitem{Mancini2015}
Evgeny~V Shchukin and Stefano Mancini.
\newblock Quantum tomography and nonlocality.
\newblock {\em Physica Scripta}, 90(7):074019, jun 2015.

\bibitem{Mancini2008}
Grigori~G. Amosov, Stefano Mancini, and Vladimir~I. Man'ko.
\newblock On the information completeness of quantum tomograms.
\newblock {\em Physics Letters A}, 372(16):2820--2824, 2008.

\bibitem{Wootters82}
W.~H.~Zurek W.~K.~Wootters.
\newblock A single quantum cannot be cloned.
\newblock {\em Nature}, 299(5886):802 -- 803, 1982.

	\bibitem{Weigert2006}
Stefan Weigert.
\newblock Simple minimal informationally complete measurements for qudits.
\newblock {\em International Journal of Modern Physics B},
20(11n13):1942--1955, 2006.

\bibitem{Glaser2020}
B\UseTextAccent{OT1}{\'}{a}lint Koczor, Robert Zeier, and Steffen~J. Glaser.
\newblock Fast computation of spherical phase-space functions of quantum
many-body states.
\newblock {\em Phys. Rev. A}, 102:062421, Dec 2020.

\bibitem{Baumgartner_2007}
Bernhard Baumgartner, Beatrix Hiesmayr, and Heide Narnhofer.
\newblock A special simplex in the state space for entangled qudits.
\newblock {\em Journal of Physics A: Mathematical and Theoretical},
40(28):7919, jun 2007.

\bibitem{Vogel1989}
K.~Vogel and H.~Risken.
\newblock Determination of quasiprobability distributions in terms of
probability distributions for the rotated quadrature phase.
\newblock {\em Phys. Rev. A}, 40:2847--2849, Sep 1989.

\bibitem{Raymer1993}
D.~T. Smithey, M.~Beck, M.~G. Raymer, and A.~Faridani.
\newblock Measurement of the wigner distribution and the density matrix of a
light mode using optical homodyne tomography: Application to squeezed states
and the vacuum.
\newblock {\em Phys. Rev. Lett.}, 70:1244--1247, Mar 1993.

\bibitem{Mancini1996}
S.~Mancini, V.I. Man'ko, and P.~Tombesi.
\newblock Symplectic tomography as classical approach to quantum systems.
\newblock {\em Physics Letters A}, 213(1):1--6, 1996.

\bibitem{Munro2002}
R.~T. Thew, K.~Nemoto, A.~G. White, and W.~J. Munro.
\newblock Qudit quantum-state tomography.
\newblock {\em Phys. Rev. A}, 66:012303, Jul 2002.


\bibitem{Lueders1951}
Gerhart L\"{u}ders.
\newblock \"{U}ber die zustands\"{a}derung durch den me{$\beta$}proze{$\beta$}.
\newblock {\em Annalen der Physik}, pages 322--328, 1951.

\bibitem{Ma2014}
Teng Ma, Ming-Jing Zhao, Yao-Kun Wang, and Shao-Ming Fei.
\newblock Non-commutativity and local indistinguishability of quantum states.
\newblock {\em Scientific Reports}, 4(1):6336, Sep 2014.

\bibitem{Walgate_2008}
Jonathan Walgate and A~J Scott.
\newblock Generic local distinguishability and completely entangled subspaces.
\newblock {\em Journal of Physics A: Mathematical and Theoretical},
41(37):375305, aug 2008.

\bibitem{Walgate2000}
Jonathan Walgate, Anthony~J. Short, Lucien Hardy, and Vlatko Vedral.
\newblock Local distinguishability of multipartite orthogonal quantum states.
\newblock {\em Phys. Rev. Lett.}, 85:4972--4975, Dec 2000.

\end{thebibliography}
\end{document}